\documentclass[cernpreprint, atlasdraft=false, texlive=2023, UKenglish, coverpage=false, texmf, orcidlogo]{atlasdoc}
 
\usepackage{atlaspackage}
\usepackage{atlasbiblatex}
 
\usepackage{atlasphysics}

\graphicspath{{logos/}{figures/}}
 
\usepackage{multirow}
 
\usepackage{ANA-HIGG-2020-07-PAPER-defs}

\AtlasTitle{
Measurement of the Higgs boson mass in the \ensuremath{H \rightarrow ZZ^* \rightarrow 4\ell} decay channel
using 139~\mbox{fb\(^{-1}\)} of $\sqrt{s}=13$~TeV $pp$ collisions recorded by the ATLAS detector at the LHC
}
 
\AtlasJournalRef{Phys. Lett. B 843 (2023) 137880}
\AtlasDOI{10.1016/j.physletb.2023.137880}
\AtlasAbstract{
The mass of the Higgs boson is measured in
the \ensuremath{H \rightarrow ZZ^* \rightarrow 4\ell} decay channel.
The analysis uses proton--proton collision data from the Large Hadron Collider
at a centre-of-mass energy of 13~TeV recorded by the ATLAS detector between 2015 and 2018,
corresponding to an integrated luminosity of 139~\mbox{fb\(^{-1}\)}.
The measured value of the Higgs boson mass is
$124.99\pm0.18\text{(stat.)}\pm0.04\text{(syst.)}$~GeV.
In final states with muons, this measurement benefits from an improved momentum-scale
calibration relative to that adopted in previous publications.
The measurement also employs an analytic model that takes into account
the invariant-mass resolution of the four-lepton system on a per-event basis and the output of
a deep neural network discriminating signal from background events.
This measurement is combined with the corresponding measurement using 7 and 8~TeV $pp$\
collision data, resulting in a Higgs boson mass of $124.94\pm0.17\text{(stat.)}\pm0.03\text{(syst.)}$~GeV.
}
 
\author{The ATLAS Collaboration}
 
\PreprintIdNumber{CERN-EP-2022-135}
 
\AtlasRefCode{HIGG-2020-07}

\hypersetup{pdftitle={ATLAS document},pdfauthor={The ATLAS Collaboration}}
 
\begin{document}
 
\maketitle
 
\tableofcontents

\section{Introduction}
 
The observation of a Higgs boson, $H$, in 2012 by the ATLAS and CMS
collaborations~\cite{HIGG-2012-27, CMS-HIG-12-028} in proton--proton ($pp$)
collisions produced by the Large Hadron Collider (LHC)
was a major step towards understanding the mechanism of the electroweak (EW) symmetry
breaking~\cite{Englert:1964et,Higgs:1964pj,Guralnik:1964eu}.
An unknown parameter of the Standard Model (SM), the Higgs boson mass \mH
is related to the SM vacuum stability~\cite{SHER1989273}\  and its value is required for precise calculations
of EW observables, including the production and decay properties of the Higgs boson itself.
These calculations are needed to test the coupling structure of the SM Higgs boson,
as suggested in Ref.~\cite{deFlorian:2016spz} and references therein.
Therefore, it is important to determine the Higgs boson mass experimentally.
 
The mass of the Higgs boson was measured to be $125.09 \pm 0.24~\GeV$~\cite{HIGG-2014-14}
in a combined analysis performed by the ATLAS and CMS collaborations
using approximately 25~\ifb\ of $\sqrt{s}=7$ and 8~\tev\ $pp$ collision data recorded in 2011 and 2012, respectively, commonly referred to as Run 1.
The individual measurements, reported in Refs.~\cite{HIGG-2013-12,CMS-HIG-14-009},
used the $\hfourl$\ (where $\ell = e$ or $\mu$) and \hgg\ decay modes because of their excellent mass resolution.
 
The ATLAS Collaboration reported a measurement of \mH\ using
the $\hfourl$\ and \hgg\ channels with 36.1~\ifb\ of $\sqrt{s}=13$~\tev\ $pp$ collision data
recorded in 2015 and 2016.
This was combined with the Run~1 ATLAS measurement to obtain a  value of
$\mH=124.97 \pm 0.24$~\GeV~\cite{HIGG-2016-33}.
The CMS Collaboration measured \mH\ using
the $\hfourl$\ and \hgg\ channels with 35.9~\ifb\ of $\sqrt{s}=13$~\tev\ $pp$ collision data
recorded in 2016.
This was combined with the Run~1 CMS measurement to obtain a  value of
$\mH=125.38 \pm 0.14$~\GeV~\cite{CMS-HIG-19-004}.
 
This paper reports a new measurement of \mH\ in the \hfourl\ channel
using a \intlumi~\ifb\ dataset of 13~\TeV\ $pp$\ collisions produced by the LHC and
recorded by the ATLAS detector between 2015 and 2018, commonly referred to as Run 2.
Most of the techniques used in this analysis are described in detail in Ref.~\cite{HIGG-2018-28}.
 
The \hfourl\ process is identified by selecting four leptons ($\ell=e,\mu$) in the final state, and
by identifying one pair of same-flavour leptons that is consistent with arising from an on-shell Z-boson decay.
The mass measurement is performed using an
analytic model to describe the distribution of the reconstructed \fourl\ invariant mass,
\mfourl, as a function of \mH\ and to fit to the observed distribution.
In comparison with the previous ATLAS result from the \hfourl\ channel~\cite{HIGG-2016-33},
the measurement has been improved by using the complete Run~2 dataset
and a new high-precision muon momentum calibration.
The \mH\ uncertainty has also been reduced by introducing a neural-network-based classifier to discriminate between signal and background processes and by including the event-by-event
invariant-mass resolution of the four-lepton system in the analytic model used to fit the collision data.
 
This paper is organised as follows.
The ATLAS detector is described in \Section{sec:detector}, and the
data samples and Monte Carlo simulations are presented in \Section{sec:samples}.
The reconstruction of muons and electrons, as well as their calibration procedures, is discussed in \Section{sec:objects},
while the event selection is explained in \Section{sec:eventselection}.
The statistical models describing the signal and background samples are defined in
\Section{sec:model_fourl}\ and results of the measurement are presented in
\Section{sec:Results_4l}.
The conclusions are provided in \Section{sec:conclusion}.


\section{ATLAS detector}
\label{sec:detector}
The ATLAS experiment \cite{PERF-2007-01}  at the LHC is a multi-purpose particle detector
with nearly $4\pi$ coverage in solid
angle.\footnote{\AtlasCoordFootnote}  It includes an inner tracking detector (ID)
used for charged-particle tracking, surrounded by a 2~T solenoid.
Electromagnetic
(EM) and hadronic calorimeters are placed outside the solenoid, followed by
a muon spectrometer (MS).
 
The ID provides precise reconstruction of tracks within a
pseudorapidity range $|\eta| \leq 2.5$.  The high-granularity silicon
pixel detector covers the vertex region and typically provides four
measurements per track, the first hit normally being in the insertable
B-layer (IBL) installed before
Run~2~\cite{ATLAS-TDR-2010-19,PIX-2018-001}.  It is followed by the
silicon microstrip tracker, which usually provides eight
measurements per track.  These silicon detectors are complemented by
the transition radiation tracker (TRT), which enables radially
extended track reconstruction up to \(|\eta| = 2.0\).
The TRT also provides electron-identification information based
on the detection of transition radiation X-ray photons.
 
The ATLAS calorimeter system
covers the pseudorapidity range $|\eta|<4.9$, with
finer granularity over the region matching the inner detector.
The lead/liquid-argon (LAr) EM calorimeter is divided into two half-barrels ($|\eta|<1.475$) and two endcap components ($1.375<|\eta|<3.2$).
It is segmented into three longitudinal (depth) sections over the region $|\eta|<2.5$,
and into two depth sections for $2.5<|\eta|<3.2$.
Hadronic calorimetry is provided by the steel/scintillator-tile
calorimeter, segmented into three barrel structures within \(|\eta| <
1.7\), and two copper/LAr hadronic endcap calorimeters extending the coverage to $|\eta| = 3.2$.
The solid angle coverage is completed with forward copper/LAr and tungsten/LAr calorimeter modules
optimised for electromagnetic and hadronic energy measurements, respectively.
 
The muon spectrometer, located beyond the calorimeters, is designed to detect muons
in the region $|\eta| < 2.7$\ and to provide momentum
measurements with a relative resolution better than 3\% over a wide
range of transverse momentum, \pT, and up to 10\% at \pT $\sim$ 1 \TeV.
A system of three superconducting air-core toroidal magnets
provides a magnetic field with a bending power of approximately
2.5 Tm in the barrel and up to 6 Tm in the
endcaps.
 
A two-level trigger system~\cite{TRIG-2016-01} is used to select
events. The first-level trigger is implemented in hardware and uses a
subset of the detector information to accept events at a maximum rate
of about \SI{100}{\kHz}.  This is followed by a software-based
trigger that reduces the accepted event rate to \SI{1}{\kHz} on
average.
An extensive software suite~\cite{ATL-SOFT-PUB-2021-001} is used
in the reconstruction and analysis of real and simulated data, in detector operations,
and in the trigger and data acquisition systems of the experiment.


\section{Data and event simulation}
\label{sec:samples}
 
This measurement uses data from $pp$ collisions with a centre-of-mass
energy of $13~\TeV$ collected between 2015 and 2018 using single-lepton,
dilepton, and trilepton triggers~\cite{TRIG-2018-01,TRIG-2018-05} as detailed in Ref.~\cite{HIGG-2018-28}.
The combined efficiency of these triggers is approximately 98\%, 99\%, 97\%, and 99\% for the
$4\mu$, $2\mu2e$, $2e2\mu$, and $4e$ final states, respectively, for the simulated \hfourl\ events
passing the event selection described in \Section{sec:eventselection}\
(assuming $\mH=125$~\GeV).
After data-quality requirements are imposed,
the integrated luminosity of the data sample is \intlumi~\ifb~\cite{DAPR-2018-01}.
 
Details of the simulated Monte Carlo (MC) events used in this analysis can be found
in Refs.~\cite{HIGG-2018-28,HIGG-2018-29}.
Higgs boson production via the gluon--gluon fusion (ggF) process was modelled at next-to-next-to-leading-order (NLO) accuracy
in the strong coupling constant \alphas\ using the \POWHEG\ NNLOPS generator~\cite{
Hamilton:2013fea,
Hamilton:2015nsa,
Alioli:2010xd,
Nason:2004rx,
Frixione:2007vw,
Hamilton:2012np,
Campbell:2012am,
Hamilton:2012rf,
Catani:2007vq}
with the \PDFforLHC[15nnlo] set of parton distribution functions (PDFs)~\cite{Butterworth:2015oua}.
Higgs bosons produced via vector-boson fusion (\VBF),
in association with a vector boson (\VH) or in association with a top-quark pair (\ttH),
were simulated at NLO accuracy with the
\POWHEGBOX generator~\cite{Alioli:2010xd,Frixione:2007vw,Nason:2004rx},
using the \PDFforLHC[15nlo] PDF set.
The loop-induced $gg \rightarrow ZH$\ process was simulated with the \POWHEGBOX generator at leading-order (LO) accuracy.
Higgs boson production in association with a top quark (\tH)
or with a bottom-quark pair (\bbH)
was simulated at NLO accuracy using the \MGNLO{} generator~\cite{Alwall:2014hca,Wiesemann:2014ioa} with the
\NNPDF[3.0]\ PDF set~\cite{Ball:2014uwa}.
For all signal processes, the \EVTGEN[1.2.0] generator~\cite{Lange:2001uf} was used for the simulation of the
bottom- and charm-hadron decays.
Correspondingly, the \PYTHIA[8] generator~\cite{Sjostrand:2007gs}\ was used for
the \htollllbrief{} decay as well as for parton showering, hadronisation, and simulation of the underlying event.
Higgs bosons produced via the \ggF\ and \VBF processes or in association with a vector boson were simulated for a range of \mH\
values from 123 to 127~\GeV.
Higgs boson production in association with a top quark, a top-quark pair,
or a bottom-quark pair were simulated assuming $\mH=125$~\GeV, as these processes make a negligible contribution.
These samples are normalised to cross-sections obtained from the most recent predictions provided by the LHC Higgs Working Group~\cite{deFlorian:2016spz}.
 
The \zzstar\ continuum background was modelled separately for
quark-initiated ($qqZZ$),
gluon-initiated ($ggZZ$),
and vector boson scattering (EW $ZZ$) production.
The $qqZZ$\ process was modelled at NLO accuracy in \alphas\ using the \Sherpa[2.2.2]
generator~\cite{Gleisberg:2008ta,Gleisberg:2008fv,Cascioli:2011va,Bothmann:2019yzt}
with the \NNPDF[3.0nnlo] PDF set.
The $ggZZ$\ process with 0 or 1~jet in the final state was modelled with a LO calculation
using the same Sherpa generator setup, with higher-order corrections calculated
using massless quark loops~\cite{Caola:2015psa,Caola:2015rqy,Campbell:2016ivq} in the
heavy top-quark approximation~\cite{Melnikov:2015laa}.
The resulting $ggZZ$\ cross-section was scaled by $1.7\pm1.0$\ to account for the
uncertainty in additional higher-order effects.
Production of EW~$ZZ$ was modelled at LO accuracy with the
\Sherpa[v2.2.2]\ with matching of the matrix element to the parton shower using the
ME+PS@NLO prescription~\cite{Hoeche:2012yf}.
These samples were normalised to cross-sections obtained directly from the \Sherpa\ simulation.
Alternative $qqZZ$ samples,
produced with the \POWHEGBOX\ v2 and \MGNLO\ generators,
are used for studies of the systematic uncertainties.
For the nominal and alternative \zzstar\ continuum samples, the \PYTHIA[8] generator was used for parton showering,
hadronisation, and simulation of the underlying event.
 
Backgrounds from $WZ$ production and $t\bar{t}$ events were modelled using the \POWHEGBOX[v2] generator,
while $Z$ bosons produced in association with jets were simulated using the \Sherpa[2.2.1]
generator.
Minor contributions from processes with three electroweak bosons, denoted by $VVV$,
were modelled using \Sherpa[2.2.2]. Small backgrounds originating from top-quark production in
association with one or more electroweak bosons or additional top quarks, such as \ttZ, $tWZ$, \ttWW,
\ttWZ, \ttZg, \ttZZ, \ttt, \tttt, and $tZ$\ (denoted by $tXX$),
were simulated using the \MGNLO\ generator. The \PYTHIA[8] generator was used for
parton showering, hadronisation, and simulation of the underlying event.
 
Generated events were processed through the ATLAS detector simulation~\cite{SOFT-2010-01}
within the \GEANT framework~\cite{Agostinelli:2002hh} and
reconstructed in the same way as collision data.
Additional $pp$ interactions in the same or neighbouring bunch crossings, referred
to as \pileup, are included in the simulation.  The \pileup\ events were modelled
using the \PYTHIA[8] generator with the A3 set of tuned parameters~\cite{ATL-PHYS-PUB-2016-017}
and the \NNPDF[2.3lo] PDF set~\cite{Ball:2012cx}.
 
The normalisation of the non-resonant \zzstar\ background component
is determined by the fit of a signal and background model to the observed \mfourl\ distributions.
The normalisations of the backgrounds from hadrons or hadron-decay products misidentified as prompt leptons,
and from $Z$+jets, $t\bar{t}$ and $WZ$ processes (referred to as the reducible background),
are determined using data-driven techniques, and are explained in detail in Ref.~\cite{HIGG-2018-28}.


\section{Muon and electron reconstruction}
\label{sec:objects}
 
Muon candidates are reconstructed using
a combination of different algorithms~\cite{MUON-2018-03}.
The reconstruction of muon candidates within $|\eta|<2.5$ is primarily
performed by a global fit of reconstructed tracks in the ID and the MS.
In the central detector region ($|\eta|<0.1$),
where the MS has reduced geometrical coverage, muons are also identified
by matching a reconstructed ID track to either an MS track
segment (`segment-tagged muons') or a calorimetric energy deposit
consistent with a minimum-ionising particle (`calorimeter-tagged muons').
Calorimeter-tagged muons are required to have $\pt>15$~\GeV.
For both the segment-tagged and calorimeter-tagged muons, the muon momentum
is measured from the ID track alone.
In the forward MS region ($2.5<|\eta|<2.7$)
outside the ID coverage, MS tracks with hits in three MS layers
are accepted as `stand-alone muons' and combined with track segments formed from
hits in the silicon tracker, if they exist.
Additionally, `loose' muon-identification criteria~\cite{MUON-2018-03} are applied to reject
low-quality tracks that have missing hits in the MS or have poor
agreement between the reconstructed MS and ID tracks.
These requirements have an efficiency of at least $98\%$ for muons with \pt\ above 5~\GeV.
Muons are required to be isolated by using
both calorimeter-based and track-based isolation variables and applying the `PflowLoose' criteria~\cite{MUON-2018-03}.
The efficiency of the isolation selection is about $85\%$ for muons with \pt\ above
5~\GeV and increases to ${>}97\%$ for muons with $\pt>20$~\GeV.
 
Although the detector is well aligned overall, there are residual local misalignments that effect
the reconstructed muon-track sagitta~\cite{IDTR-2019-05,ATL-MUON-PUB-2021-002} and
introduce a small, charge-dependent, momentum bias.
The bias is measured with an iterative procedure that
minimises the observed width of the line-shape of \Zmumu\ decays~\cite{MUON-2022-01}.
This correction, which is applied to the collision data,
reduces the width of the dimuon invariant-mass distribution in $Z$-boson decays by approximately 1\%~\cite{MUON-2022-01}.
A systematic uncertainty is assigned to this correction to cover
residual differences between observed and simulated data.
It is estimated to be, on average, 1\% of the correction itself for muons from \Zmumu\ decays.
 
Corrections are also applied to the reconstructed muon momentum in simulation in order to
precisely match the data.
These corrections to the simulated momentum resolution and momentum scale are parameterised as
an expansion in powers of the muon \pt, with each coefficient measured as a function of $\eta$ and $\phi$.
The corrections are extracted from large data samples of \Jpsimumu\ and \Zmumu\ decays,
using techniques that reduce any residual biases
and by directly calibrating the global re-fitted track~\cite{MUON-2022-01}.
Compared to the previously used corrections, this reduces the associated uncertainties by approximately a factor of four.
 
The momentum-scale corrections range from 0.1\% to 0.3\% for muons with \pt between 5 and 100~\GeV.
These account for inaccuracies in the estimation of the energy loss in the traversed material,
local magnetic field inaccuracies, and geometrical distortions. The
corrections to the momentum resolution for
muons with \pt of 5--100~\GeV\
are at the percent level.
The main sources of systematic uncertainty for these corrections
are the residual biases in the calibration method and the consistency of calibrations using
\Jpsimumu\ or \Zmumu\ decays separately.
For muons from \Zmumu\ decays, which have an average \pt of approximately 45~\GeV, the momentum scale is determined with a precision better than $0.01\%$.
The resolution is known with a precision of $0.4\%$ for
muons with $|\eta|<1$, while for muons in high-$|\eta|$ regions the precision is approximately~1\%.
 
A reconstructed electron consists of a cluster of energy deposits in the calorimeter and a matched ID track~\cite{EGAM-2018-01}.
Variable-size clusters are created dynamically from
calorimeter-energy deposits, improving the invariant-mass resolution of the four-lepton system,
especially when bremsstrahlung photons are present.
Electron ID tracks are fitted using an
optimised Gaussian-sum filter (GSF)~\cite{PERF-2017-01} that accounts
for non-linear effects arising from energy loss through bremsstrahlung.
Quality criteria are used to improve the purity of selected electron candidates.
The quality of an electron candidate is evaluated by using a likelihood method
that employs measurements from the tracking system and the calorimeter system,
and quantities that combine both tracking and calorimeter information~\cite{PERF-2017-01}.
The `loose' likelihood criteria, together with track hit requirements, are applied to electron candidates.
Electrons are required to be isolated using
both the calorimeter-based and track-based isolation variables as discussed in Ref.~\cite{HIGG-2018-28}.
 
The energy of electrons is estimated from the calorimeter energy clusters,
using a combination of simulation-based and data-driven corrections~\cite{PERF-2017-03,EGAM-2018-01}.
A single simulation-based correction, which accounts for the energy
lost in the material upstream of the calorimeter,
the energy deposited in the cells neighbouring the cluster in $\eta$ and $\phi$,
and the energy lost beyond the LAr calorimeter, is derived using multivariate
regression algorithms.
Data-driven corrections account for effects such as those associated with
the material in front of the EM calorimeter and material between
the presampler and the calorimeter, and the inter-calibration of the different calorimeter layers.
The remaining energy-scale difference between data and simulation is parameterised
by energy-independent linear corrections, defined in different regions of $\eta$.
Similarly, deviations of the energy resolution in the simulation from that in the data are
parameterised as an $\eta$-dependent additional term.
The electron reconstruction efficiency is ${>}97$\%\ for electron $\pt > 18$~\GeV
and the additional identification efficiency varies between 75 and 90\%\ for
the kinematic region relevant to this analysis.
 
The main sources of systematic uncertainties in the electron-energy
scale include uncertainties in the method used to extract the energy scale
correction, as well as uncertainties due to the extrapolation
of the energy scale from \Zee\ events to
electrons with different energies.
A detailed explanation of these uncertainties can be found in Ref.~\cite{EGAM-2018-01}.
In the case of electrons with $\ET$ around 40~\GeV, the total relative uncertainties range
between 0.04\% and 0.2\% for most of the detector acceptance.  For
electrons with $\ET$ around 10~\GeV\ the relative uncertainty ranges between
0.3\% and 0.8\%.
 
Systematic uncertainties in the calorimeter-energy resolution for electrons arise
from uncertainties in the modelling of the sampling term
and in the measurement of the constant term in $Z$-boson decays, in
the amount of material in front of the calorimeter, and in the modelling of the
contribution to the resolution due to fluctuations in the pile-up from
additional $pp$ interactions in the same or neighbouring
bunch crossings. The uncertainty in the energy resolution for
electrons with transverse energy between 30 and
60~\GeV\ varies between 5\% and 10\%.


\section{Event selection}
\label{sec:eventselection}
 
Events are required to contain at least four isolated leptons
emerging from a common vertex and forming two pairs of
oppositely charged same-flavour leptons.  Electrons are required to be
within the geometrical acceptance of the inner detector
($|\eta|<2.47$) and to have $\et>7~\gev$, while muons
must be within the geometrical acceptance of the muon
spectrometer ($|\eta|<2.7$) and have $\pt>5~\gev$
(except for calorimeter-tagged muons, as explained in Section~\ref{sec:objects}).
At most one calorimeter-tagged or stand-alone muon is allowed per Higgs boson candidate.
The three higher-($\pt$ or $\et$) leptons in each quadruplet are required
to pass thresholds of 20, 15, and 10~\GeV, respectively, and have $\Delta R(\ell, \ell^\prime) > 0.1$.
These thresholds are chosen to maximize signal yield consistent with selecting
events with high trigger efficiency.
Contributions from misidentified leptons are reduced by requiring the lepton tracks to have low
transverse-impact-parameter significances and to be compatible with originating from a common vertex.
A detailed description of the event selection can be found in Ref.~\cite{HIGG-2018-28}.
 
The lepton pair with an invariant mass closest to the $Z$-boson mass
in each quadruplet is referred to as the leading dilepton, while
the remaining pair is referred to as the subleading dilepton.
The selected quadruplets are separated into four subchannels,
according to the flavour of the leading and subleading pairs.
In order of decreasing expected selection
efficiency and resolution, they are $4\mu$, $2e2\mu$, $2\mu2e$, and $4e$.
The \mfourl\ resolution is about 1.5~\GeV\ for subchannels with a subleading muon pair ($4\mu$ and $2e2\mu$)
and about 2.1~\GeV\ for subchannels with a subleading electron pair ($2\mu2e$ and $4e$).
Only one quadruplet is selected from each event, based on the mass of the
leading dilepton, the final state, and, for events with additional leptons, the value of the LO matrix element,
as described in Ref.~\cite{HIGG-2018-28}.
 
Final-state radiation (FSR) photons are searched for in all  events following
the procedure described in Ref.~\cite{HIGG-2018-28}.
FSR candidates are defined as collinear if their angular separation from the nearest lepton of the quadruplet satisfies $\Delta R <0.15$, and non-collinear otherwise.
Collinear FSR candidates are considered only for muons from the leading dilepton
as the electron candidate reconstruction includes collinear FSR,
while non-collinear FSR candidates
are considered for both muons and electrons from either the leading or the subleading dilepton.
Only one FSR candidate is included in the quadruplet as events with more than one FSR
candidate are rare and have negligible effect on the four-lepton mass resolution.
In cases where there are more one one FSR candidate,
preference is given to collinear FSR and to the candidate with the highest $\pt$.
FSR photons are found in $4\%$ of the events and their energy is included in the mass
computation, improving the $\mfourl$ resolution by about 1\%.
 
Finally, the four-momenta of leptons in the leading pair are recomputed by performing a kinematic fit
which constrains their invariant mass to the $Z$-boson mass, as discussed in Ref.~\cite{HIGG-2013-21}.
The fit, which takes into account lepton and FSR kinematic information, their associated experimental uncertainties, and the $Z$-boson width,
improves the $\mfourl$ resolution by about $17\%$~\cite{HIGG-2013-12}.
The lepton four-momenta reconstruction uncertainties were derived from simulation and
are corrected to match those observed in data, using a large sample of \Zboson-boson decays.
 
In the mass range $115 < \mfourl<130$~\GeV, 313 candidate events are observed. The yield is in
agreement with the expectation of $321\pm14$ events, 65\% of which are
predicted to originate from signal processes, assuming $\mH=125$~\GeV.
The \mfourl\ distributions are shown in
\Figure{fig:mflcPerChannel} for each final state and in
\Figure{fig:mflc} for the inclusive final state for the mass range of 105--160~\GeV, which is used to determine the Higgs boson mass.
 
\begin{figure}[t]
\centering
\subfloat[ ]{\includegraphics[width=0.45\textwidth]{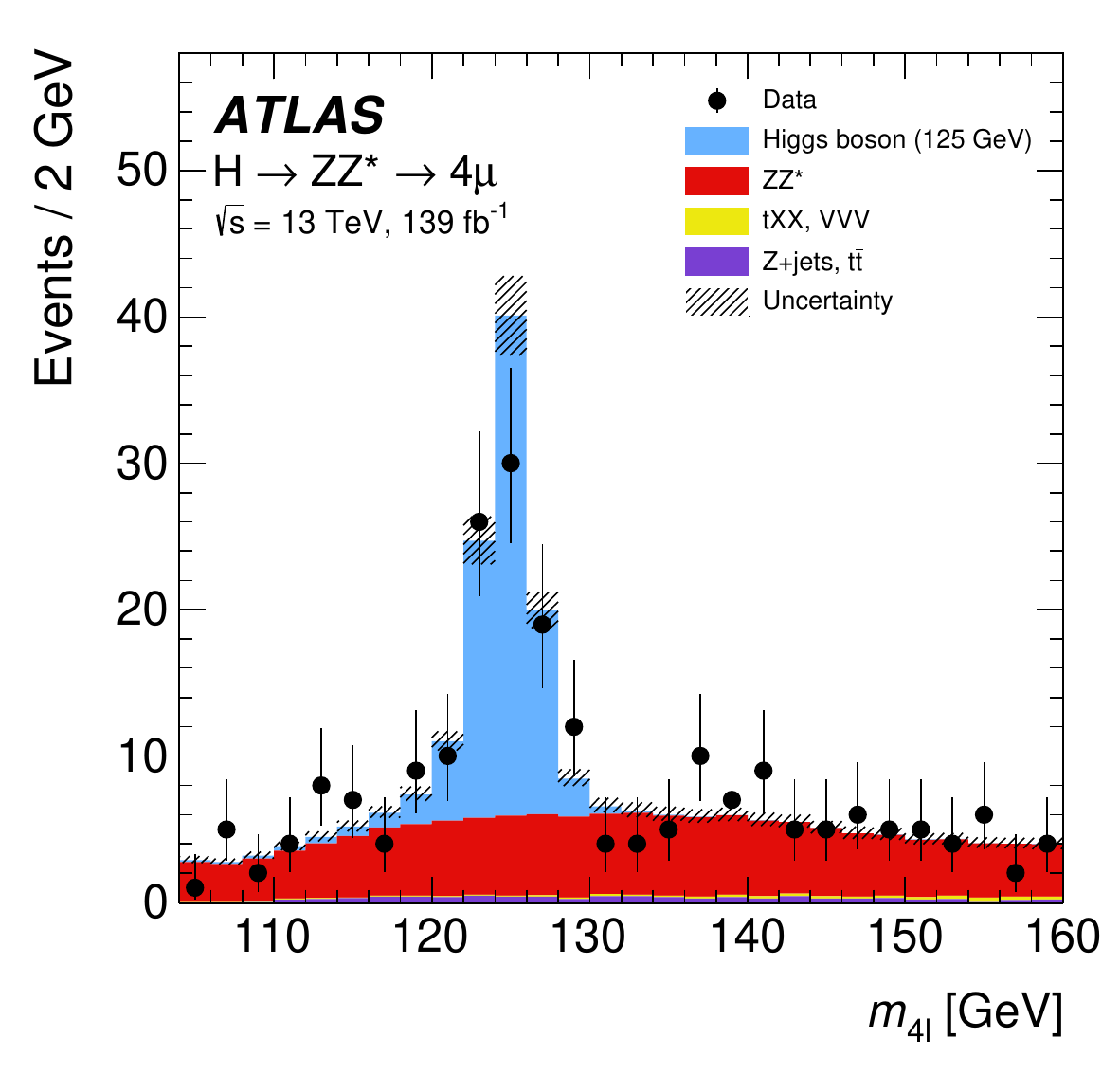}}
\subfloat[ ]{\includegraphics[width=0.45\textwidth]{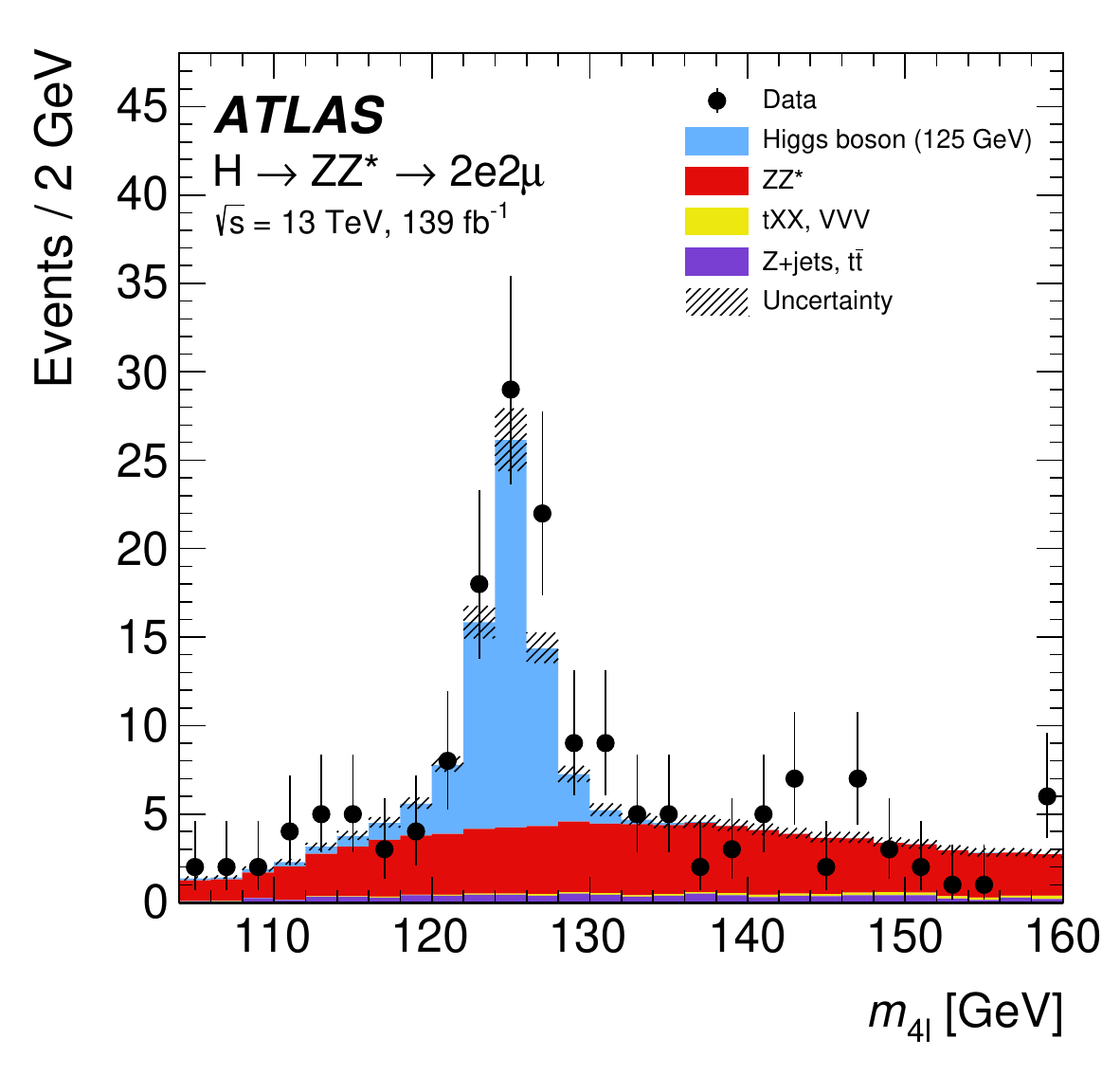}}
 
\subfloat[ ]{\includegraphics[width=0.45\textwidth]{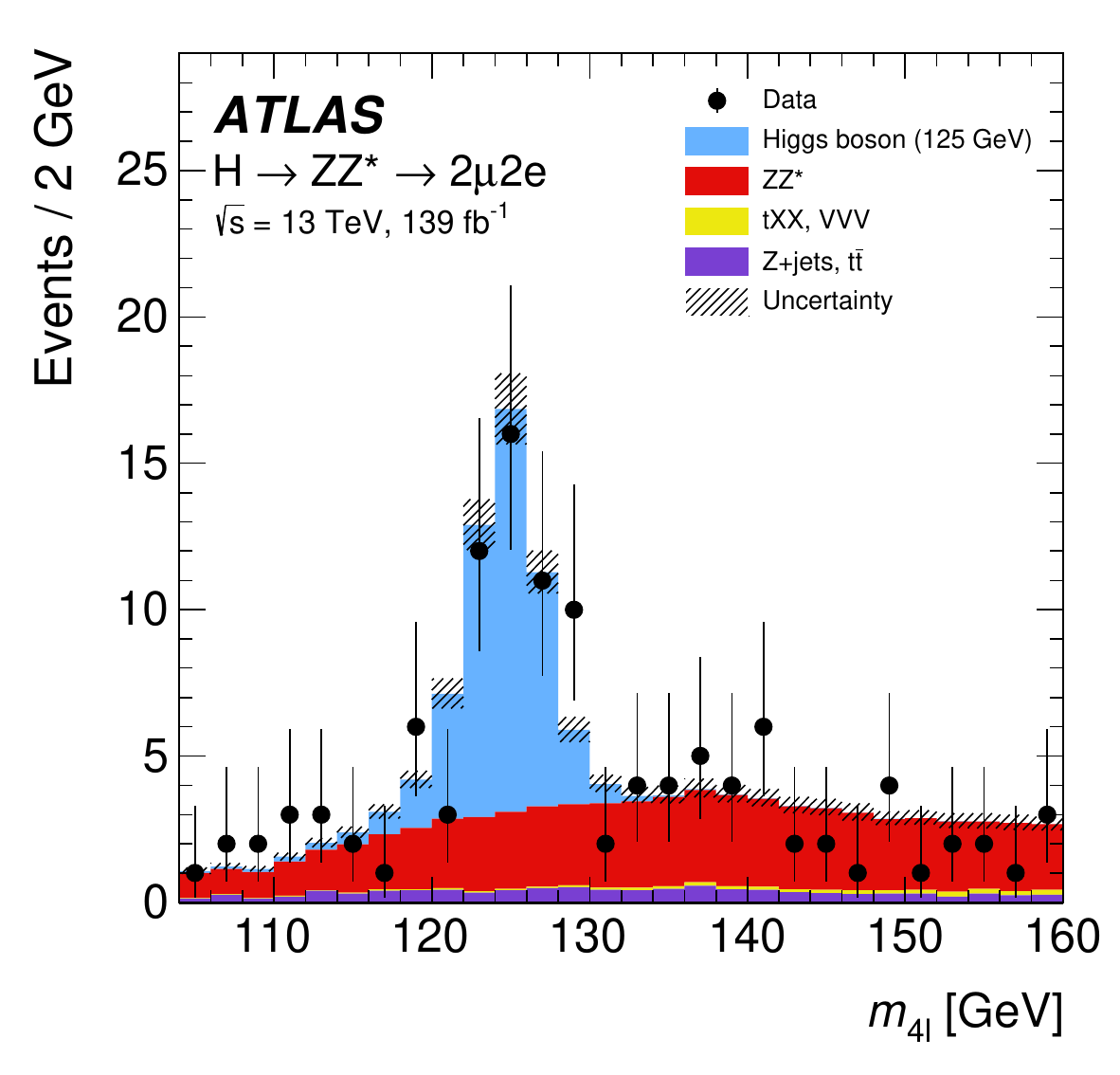}}
\subfloat[ ]{\includegraphics[width=0.45\textwidth]{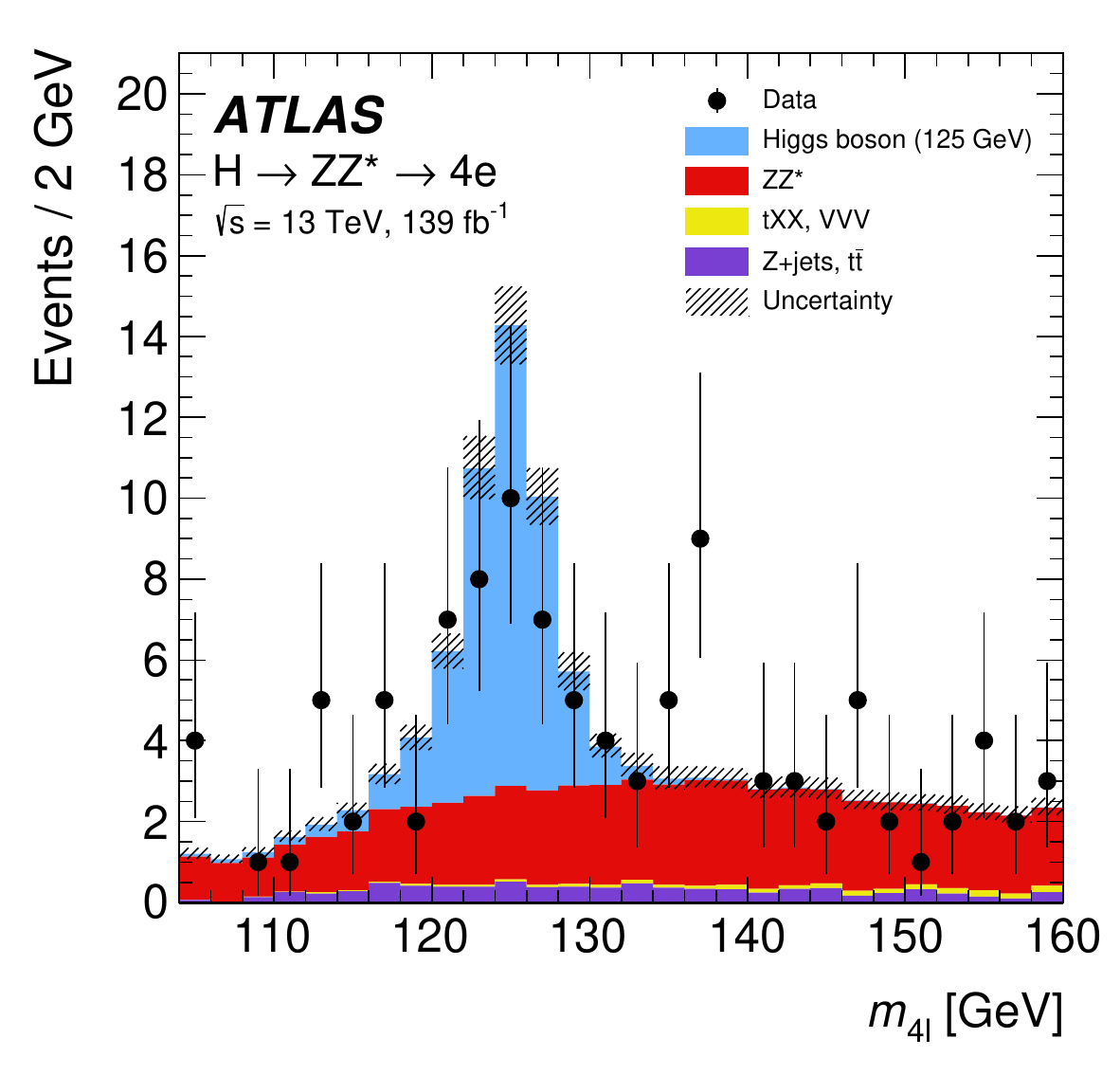}}
\caption{The observed and expected (pre-fit) \ensuremath{m_{4\ell}} distributions for the selected Higgs boson candidates, for the
different decay final states (a) $4\mu$, (b) $2e2\mu$, (c) $2\mu2e$, and (d) $4e$ events.
The predicted number of events for these distributions is taken from simulation for the signal, $ZZ^{*}$,
$tXX$, and $VVV$ processes,
while it is taken from the data-driven estimate (see \Section{sec:model_fourl}) for the $Z+$jets and \ttbar backgrounds.
The total uncertainty in the prediction is shown by the hatched band, which also includes the theoretical uncertainties of the SM cross-section for the signal and the $ZZ^{*}$ background. Higgs boson events in this plot are simulated with $\mH=125$~\GeV. The $p$-values quantifying the agreement between the data and MC predictions are (a) 0.81, (b) 0.43, (c) 0.77, and (d) 0.15.\label{fig:mflcPerChannel}}
\end{figure}
 
\begin{figure}[t]
\centering
\includegraphics[width=0.48\textwidth]{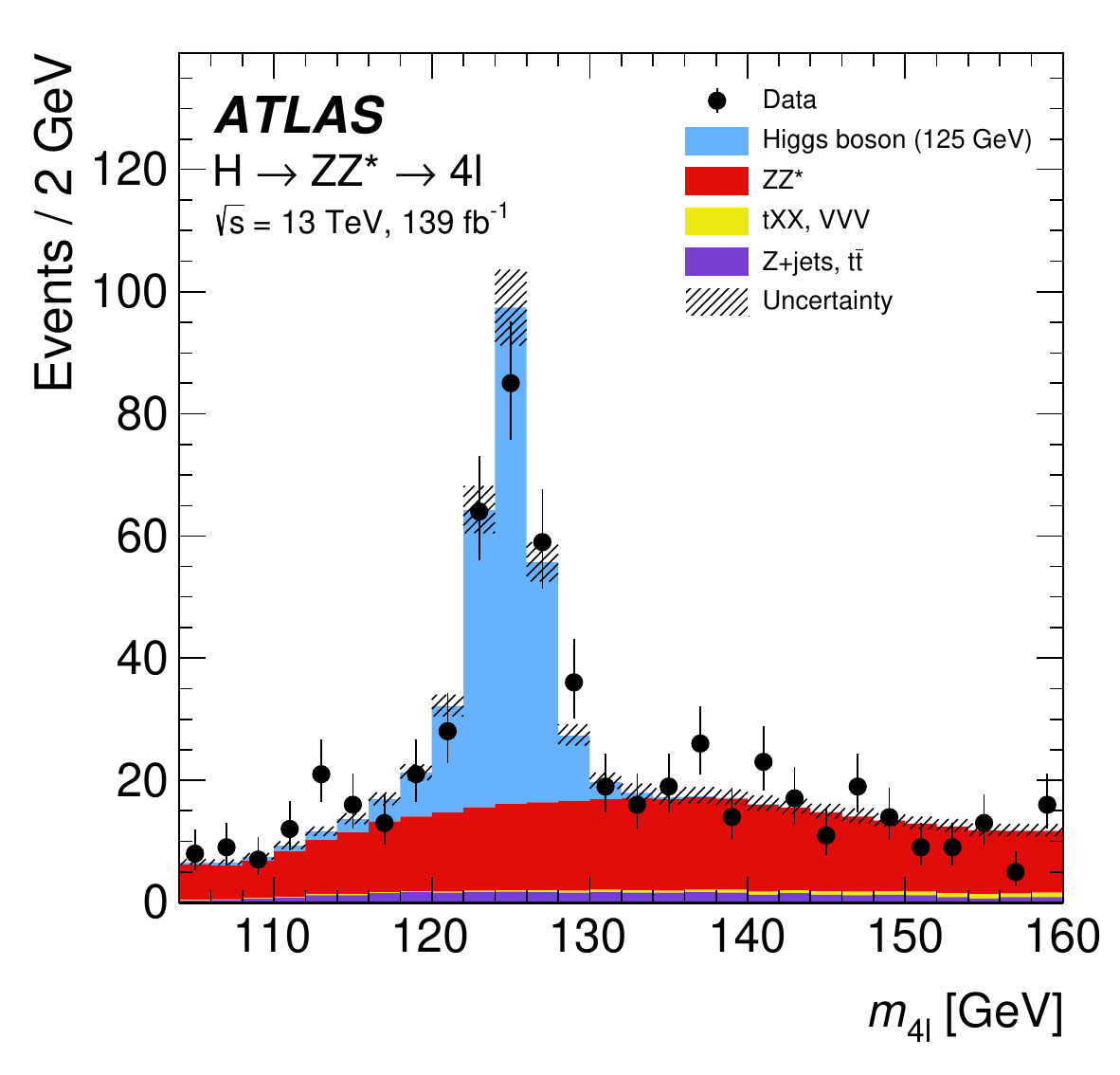}
\caption{The observed and expected (pre-fit) \ensuremath{m_{4\ell}} distributions for the selected Higgs boson candidates.
The predicted number of events for these distributions is taken from simulation for the signal, $ZZ^{*}$,
$tXX$, and $VVV$ processes,
while it is taken from the data-driven estimate (see \Section{sec:model_fourl}) for the $Z+$jets and \ttbar backgrounds.
The total uncertainty in the prediction is shown by the hatched band, which also includes the theoretical uncertainties of the SM cross-section for the signal and the $ZZ^{*}$ background. Higgs boson events in this plot are simulated with $\mH=125$~\GeV. The $p$-value quantifying the agreement between the data and MC predictions is 0.36.\label{fig:mflc}}
\end{figure}
 
The dominant contribution to the background is non-resonant
\zzstar\ production, accounting for approximately 89\% of the total background
yield. The reducible background contributes approximately 9\%.
The $VVV$ and $tXX$ events are estimated to constitute approximately 2\% of the total background. The residual
combinatorial background, originating from events with additional
prompt leptons, is found to be negligible~\cite{HIGG-2018-28}.

\subsection{Signal--background discriminant}
\label{sec:nn}
 
To provide additional separation between the $\hfourl$ signal and the $ \zzstar\to 4\ell$ background,
a deep feed-forward neural network (NN) is employed.
The NN is trained with Keras~\cite{chollet2015keras}
using the TensorFlow~\cite{tensorflow2015-whitepaper} backend,
following the method described in Ref.~\cite{HIGG-2018-28}.
Signal events for training are taken
from simulated samples with different masses of the Higgs boson, as described in Section \ref{sec:samples},
thus reducing the dependence of the discriminant on \mH.
 
The \pT\ and $\eta$\ of the four-lepton system
are used as inputs to the NN, together with
a matrix-element-based kinematic discriminant
$D_{\zzstar}$~\cite{HIGG-2013-21}.
The discriminant $D_{\zzstar}$ is defined as
$\ln(|\mathcal{M}_{H\zzstar}|^2/|\mathcal{M}_{\zzstar}|^2)$ where
$\mathcal{M}_{H\zzstar}$ denotes the matrix element for leading-order (LO) $\hfourl$ production and $\mathcal{M}_{\zzstar}$
the sum of the corresponding matrix elements for the $q\bar{q}\to \zzstar$
and $gg\to \zzstar$ continuum backgrounds, all calculated at LO with
the \MADGRAPH[2.2.1] generator~\cite{Alwall:2014hca}.
 
The distribution of the output of the NN, \nndisc, is shown in \Figure{fig:sig_bkg_disc}.
The \nndisc\ for each event is included in the final fit
as an additional observable, as explained in \Section{sec:model_fourl}.
The additional separation between signal and background provided by the NN improves
the precision of this measurement by $2\%$.
 
\begin{figure}[t]
\centering
\includegraphics[width=0.48\textwidth]{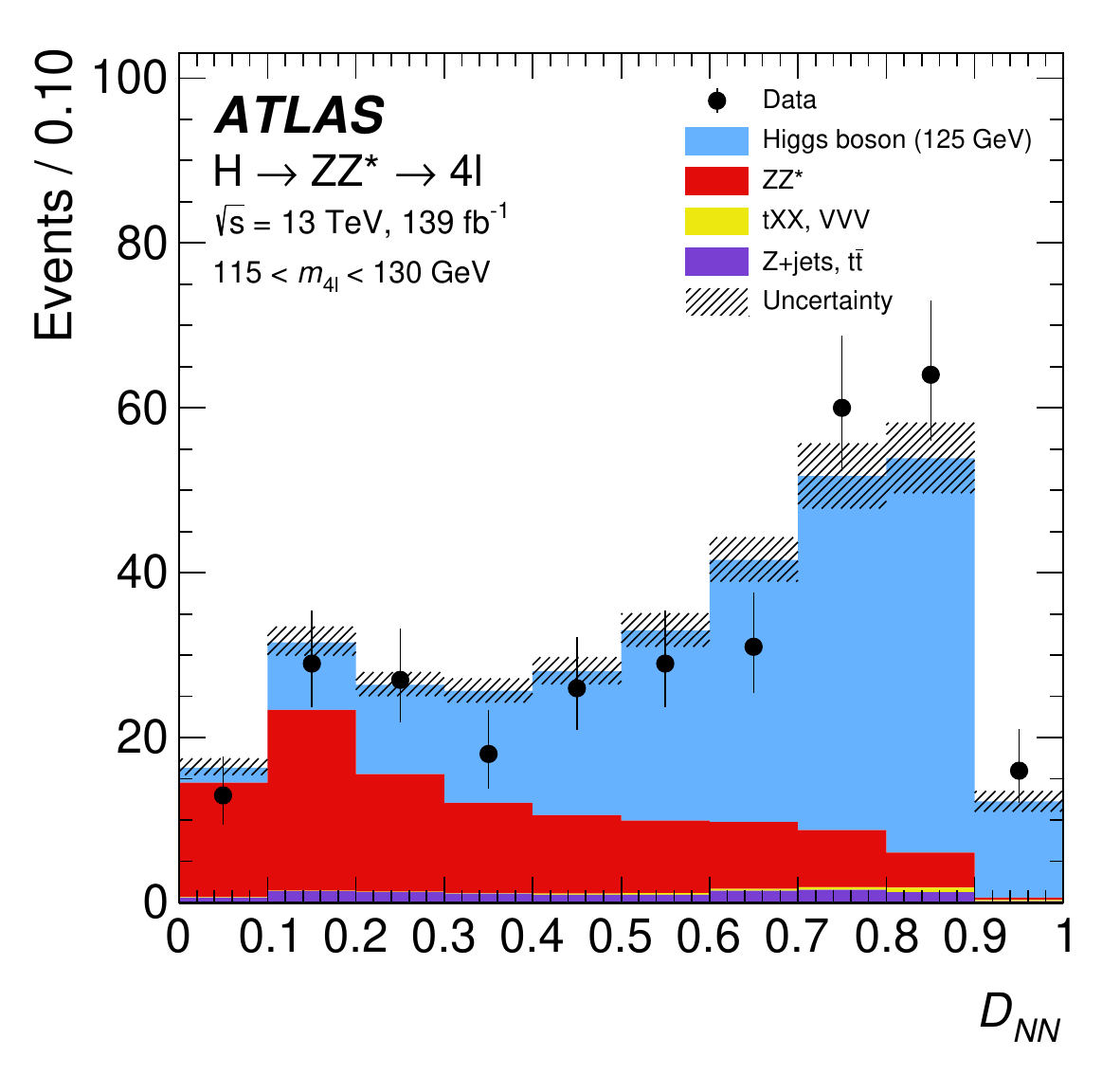}
\caption{The observed and expected (pre-fit) \nndisc\ distributions for the selected Higgs boson candidates.
The predicted number of events for these distributions is taken from simulation for the signal, $ZZ^{*}$,
$tXX$, and $VVV$ processes,
while it is taken from the data-driven estimate (see \Section{sec:model_fourl}) for the $Z+$jets and \ttbar backgrounds.
The total uncertainty in the prediction is shown by the hatched band, which also includes the theoretical uncertainties of the SM cross-section for the signal and the $ZZ^{*}$ background. Higgs boson events in this plot are simulated with $\mH=125$~\GeV. The $p$-value quantifying the agreement between the data and MC predictions is 0.28.\label{fig:sig_bkg_disc}}
\end{figure}
 
\subsection{Event-level \mfourl\ resolution}
\label{sec:pereso}
 
The event-level \mfourl\ resolution, \pesig\ ,
is estimated using a quantile regression neural network (QRNN)~\cite{qrnn}.
The QRNN is trained on signal MC events using the Tensorflow library
and the Keras Python application programming interface.
The inputs are the $\pt$, $\eta$, and $\phi$ of the individual leptons, as well as
the four-lepton momentum, constrained by the \Zboson-boson mass constraint, and its uncertainty.
The output of the QRNN for each final state is the predicted quantile of the difference between the reconstructed mass
and the true mass of the four-lepton system. The targeted quantile is calibrated using a step-wise scanning procedure so that
it results in an estimate that produces 68\%\ confidence-level intervals when tested on
simulated events.
 
The \pesig\ distributions are shown in \Figure{fig:sigma_i}.
Each event is assigned the \pesig\ estimated by the QRNN and employed in the
likelihood fit to \mH. Taking into account the per-event resolution reduces the total expected uncertainty in \mH by 1\%.
 
\begin{figure}[t]
\centering
\includegraphics[width=0.48\textwidth]{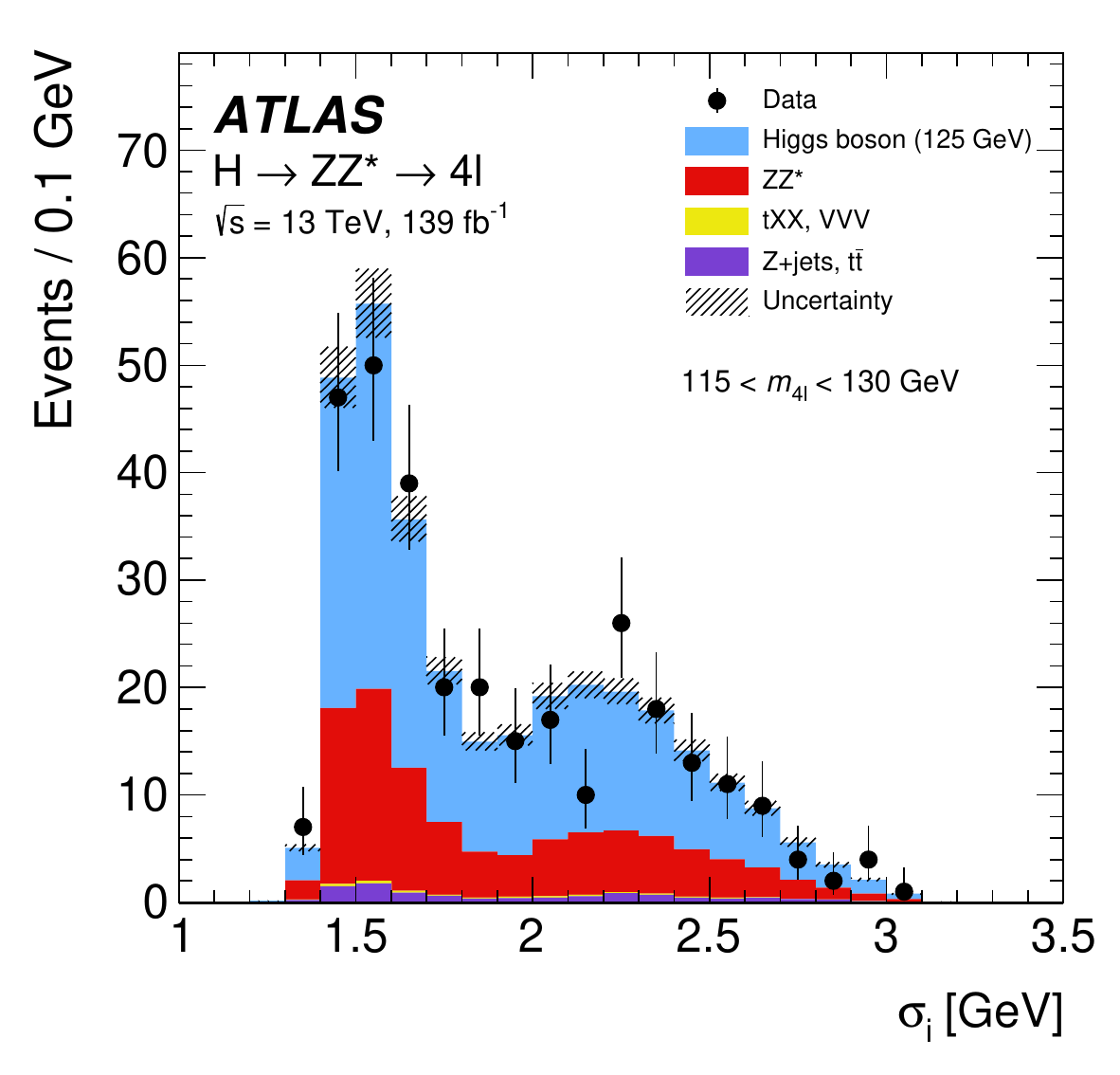}
\caption{The observed and expected (pre-fit) event-level resolution \pesig\ distributions predicted by the QRNN for the selected Higgs boson candidates.
The predicted number of events for these distributions is taken from simulation for the signal, $ZZ^{*}$,
$tXX$, and $VVV$ processes,
while it is taken from the data-driven estimate (see \Section{sec:model_fourl}) for the $Z+$jets and \ttbar backgrounds.
The total uncertainty in the prediction is shown by the hatched band, which also includes the theoretical uncertainties of
the SM cross-section for the signal and the $ZZ^{*}$ background.
Higgs boson events in this plot are simulated with $\mH=125$~\GeV.
The $4\mu$ and $2\mu2e$ events typically have $\pesig \sim 1.5$~\GeV, while
the $2e2\mu$ and $4e$ events typically have $\pesig \sim 2.1$~\GeV. The $p$-value quantifying the agreement between the data and MC predictions is 0.81. \label{fig:sigma_i}}
\end{figure}


\FloatBarrier

\section{Signal and background model} \label{sec:model_fourl}
 
The Higgs boson \mfourl\ distribution is the result of the
convolution of its theoretical line-shape, a narrow relativistic Breit--Wigner (BW) distribution
of 4.1~\MeV~width~\cite{deFlorian:2016spz} centred on
\mH, with the detector response for the four-lepton invariant-mass
distribution.  The BW width is more than two orders of magnitude smaller than
the  \mfourl\  detector resolution. Therefore, the signal line-shape in \mfourl\
is completely dominated by the detector response.
 
The signal probability density function is modelled as a function of \mfourl,
signal versus \zzstar\ discriminator \nndisc, and the per-event resolution estimator \pesig.
The model can be written as
\begin{align*}
\mathcal{P}(\mfourl,\nndisc,\pesig|\mH)
&= \mathcal{P}(\mfourl|\nndisc,\pesig,\mH) \cdot
\mathcal{P}(\nndisc|\pesig,\mH) \cdot
\mathcal{P}(\pesig|\mH) \\
&\simeq \mathcal{P}(\mfourl|\nndisc,\pesig,\mH) \cdot
\mathcal{P}(\nndisc|\mH),
\end{align*}
where the following approximations are used:
\begin{itemize}
\item $\mathcal{P}(\nndisc|\pesig,\mH) \simeq \mathcal{P}(\nndisc|\mH)$ because the neural network
discriminant does not directly depend on the per-event \mfourl\ resolution,
\item $\mathcal{P}(\pesig|\mH) \simeq \mathcal{P}(\pesig)$ since the averaged per-event resolution does not depend
on \mH within the range of 105 to 160~\GeV\ used in this measurement, and
\item $\mathcal{P}(\pesig)$ is omitted from the probability density function because it was observed that it has
approximately the same distribution for signal and background events in the Higgs boson peak region.
Dedicated checks of the assumption that $\mathcal{P}(\pesig)$ can be omitted have shown that this has negligible impact on the measurement.
\end{itemize}
 
The probability density function
$\mathcal{P}(\nndisc|\mH)$ is determined through interpolation between neighbouring \mH\ values using simulated data.
The probability density function
$\mathcal{P}(\mfourl|\nndisc,\pesig,\mH)$ in each subchannel is described by a double-sided
Crystal Ball~\cite{HIGG-2014-04}  probability density function that consists of a
Gaussian core and two power-law tails.
 
The mean of the Gaussian core is parameterised as a function of \mH and \nndisc\
for each subchannel $\lambda$ as
\begin{equation*}
a^{\lambda}\cdot(\mH-125~\GeV)+b^{\lambda}(\nndisc),
\end{equation*}
in order to decorrelate the uncertainties of the shifts in the mean reconstructed \mfourl\
due to the subchannel dependence from those due to the \nndisc\ dependence.
The $a^{\lambda}$ parameters are consistent with unity within a few percent for all final states.
The dependence of $b^{\lambda}$ on $\nndisc$ is parameterised as a
second-order polynomial, and its values lie in the 124--125~\GeV\ range for
the $2\mu2e$ and $4e$ final states,
and between $124.6$~\GeV\ and $125$~\GeV\ for the $2e2\mu$ and $4\mu$ final states.
The residual correlations between the \mH\ dependence of the mean and shape of the reconstructed \mfourl\ distribution,
and the \nndisc\ dependence of the mean have been checked with simulation and found to be negligible.
The $a^{\lambda}$ parameters and the $b^{\lambda}$ functions are extracted from an unbinned maximum-likelihood fit
to the combined \ggF, \VBF, and \VH\ samples, which were simulated with different values of \mH, as described in \Section{sec:samples}.
 
The standard deviation of the Gaussian core is expressed
as a function of the predicted event-level resolution \pesig\, and is parameterised
as a function of \nndisc\ to account for a residual correlation between \nndisc\ and \pesig.
The parameters of the left tail of the double-sided Crystal Ball are parameterised as a
function of \nndisc\ only, which results in few percent variation as a function
of \nndisc, while those
of the right tail are found to be constant for a given final state.
 
The parameters of the signal model for each subchannel are
derived by a simultaneous maximum-likelihood fit to signal MC samples
for \mH\ values ranging from 123~\GeV\ to 127~\GeV. The resulting probability density function is
shown in \Figure{fig:pdf_closure} for the $\mH=125$~\GeV\ signal sample.
The residual bias in the measured \mH\ for the derived model, estimated by pseudo-experiments
sampling MC events, is at the level of 6~\MeV\ and is treated as an additional
systematic uncertainty.
In the fit of the model to the data, \mH\ can vary freely while the other parameters are constrained to vary in
accordance with the impact of the relevant systematic uncertainties. Negligible biases in \mH\ are observed when
fitting the model to simulated Higgs signal samples with different values of \mH.

\begin{figure}[t]
\centering
\includegraphics[width=0.6\textwidth]{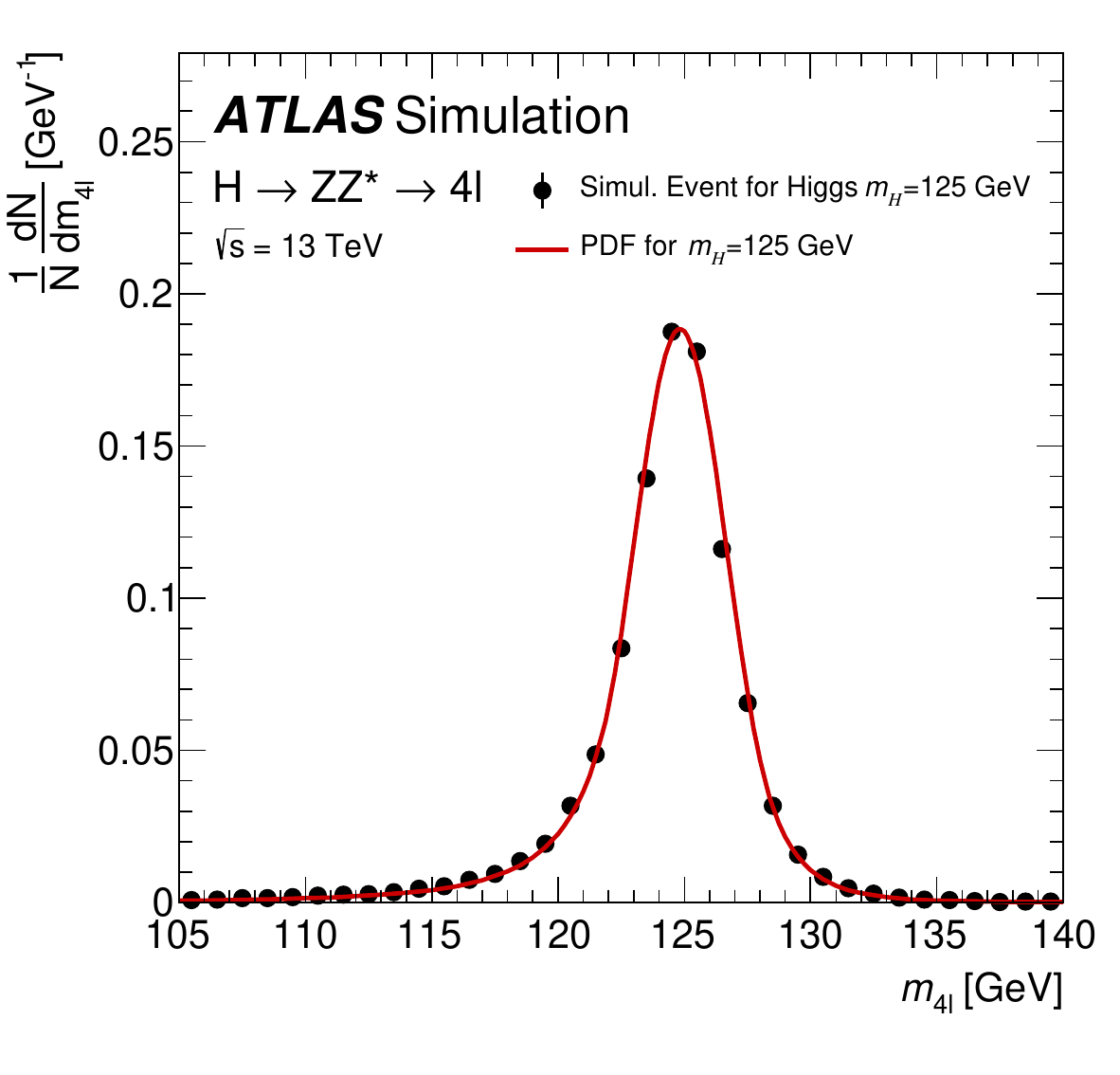}
\caption{The probability density function derived from signal MC simulation
using the per-event resolution (red line) is shown in
comparison with the
\ensuremath{m_{4\ell}} distribution in signal MC simulation
(black points) for $\mH=125$~\GeV. The probability density function
is derived for each channel separately and they are added together assuming predicted yields.
Both the probability density function and the
MC simulation are normalised to unity.
\label{fig:pdf_closure}}
\end{figure}
 
The probability density functions for the $ZZ^{*}$ and $tXX+VVV$ backgrounds are estimated
as a function of \mfourl\ and \nndisc\ for each subchannel from MC simulation.
The size of the background contributions from misidentified and non-isolated leptons are estimated
from a data-control region~\cite{HIGG-2018-28}, with the dependence on \mfourl\ and \nndisc\
parameterised using simulated samples.
 
The probability density functions for the backgrounds are smoothed using an adaptive kernel estimation technique,
employing Gaussian kernels~\cite{KEYS}\ to reduce statistical fluctuations. The effects
of uncertainties that modify the shapes of probability density functions are described by using non-linear moment morphing~\cite{Baak_2015}
to interpolate between the nominal probability density functions and those accounting for the relevant systematic variations.
Such uncertainties include the theoretical and experimental uncertainties
for the $ZZ^{*}$ and $tXX+VVV$ probability density functions and imperfect knowledge of the relative
background composition for the reducible background's probability density function.
 
The \mfourl\ distribution is described by the sum of a signal
distribution and the distributions of the backgrounds.
The normalisations of the signal contribution and the $ZZ^{*}$ process in each subchannel are free parameters
and extracted directly from the fit to data,
while those of the reducible background in each channel are constrained according to estimates
obtained from data, using minimal input from simulation and following the methodology
described in Ref.~\cite{HIGG-2018-28}. The normalisation of the $tXX+VVV$
background is constrained according to the uncertainties in the theory prediction.
 
The fit model was validated by utilising a two-step unblinding process, where the same, unknown, bias is applied
to all events in the observed dataset and the likelihood model, as described in Section~\ref{sec:Results_4l},
is fitted to it.
No significant changes to the compatibility between channels or to any free or constrained parameters, with the
exception of \mH, are observed when this bias is removed.


\FloatBarrier

\section{Results}
\label{sec:Results_4l}
 
The mass measurement is performed by maximising the profile-likelihood ratio~\cite{HIGG-2012-17,Cowan:2010js}
\begin{equation*}
\lambda(\mH) =
\frac{\Like\big(\mH,\hat{\hat{\theta}}(\mH)\big)}{\Like\big(\hat{m}_H,\hat{\theta}\big)},
\end{equation*}
where  $\hat{m}_{H}$ and  $\hat{\theta}$ denote the
unconditional maximum-likelihood estimates of the parameters of the
likelihood function \Like, while $\hat{\hat{\theta}}$ is the conditional
maximum-likelihood estimate of the parameters $\theta$ for a fixed
value of the parameter of interest, \mH.
Systematic uncertainties and their correlations are modelled by
introducing nuisance parameters $\theta$ with priors described by Gaussian or
log-normal functions that reflect the uncertainties in the values of the nuisance parameters.
 
The estimate of \mH\ is extracted by performing a simultaneous unbinned maximum-likelihood fit
to the four subchannels ($4\mu$, $2e2\mu$, $2\mu2e$, and $4e$) in the \mfourl\ range between 105 and 160~\GeV.
The expected and observed yields, along with the signal-to-background ratio,
are shown in \Table{tab:yields_signal_region} for the signal region defined by
limiting the data to \mfourl\ between 115 and 130~\GeV.
 
\begin{table}[t]
\caption{The observed and expected (pre-fit) yields for the
different decay final states in the region with \mfourl\ between 115 and 130~\GeV.
The predicted number of events is taken from simulation for the signal, $ZZ^{*}$,
$tXX$, and $VVV$ processes,
while for $Z+$jets and \ttbar it is taken from the data-driven estimate (see \Section{sec:model_fourl}).
The uncertainty in the prediction includes the theoretical uncertainties of the SM cross-section for the signal and the $ZZ^{*}$ background.
Higgs boson yields are determined using simulation with $\mH=125$~\GeV. }
\label{tab:yields_signal_region}
\centering
\begin{tabular}{ l c c c c c }
\toprule
Final & \multirow{2}{*}{Higgs}   & $ZZ$,                 & Reducible              & Expected                     & Observed \\
state &                          & $tXX$, $VVV$          & backgrounds            & total yield                  & yield    \\
\midrule
$4\mu$  & $  78    \pm  5   $      &   $  38.7  \pm  2.2 $ &   $  2.84  \pm  0.17 $ &  $  120   \pm  5\phantom{0}$ & 115~~    \\
$2e2\mu$  & $  53.4  \pm  3.2 $      &   $  26.7  \pm  1.4 $ &   $  3.02  \pm  0.19 $ &  $  83.1  \pm  3.5 $         & 94       \\
$2\mu2e$  & $  41.2  \pm  3.0 $      &   $  17.9  \pm  1.3 $ &   $  3.4   \pm  0.5  $ &  $  62.5  \pm  3.3 $         & 59       \\
$4e$  & $  36.2  \pm  2.7 $      &   $  15.7  \pm  1.6 $ &   $  2.83  \pm  0.35 $ &  $  54.8  \pm  3.2 $         & 45       \\
\midrule
Total  & $  209   \pm  13  $      &   $  99    \pm  6   $ &   $  12.2  \pm  0.9  $ &  $  321   \pm  14  $         & 313~     \\
 
\bottomrule
\end{tabular}
\end{table}
 
The free parameters of the fit are \mH, the signal normalisation, and background normalisations for each of the four
subchannels, while the nuisance parameters associated with the systematic uncertainties are constrained by their
respective priors.
\Figure{fig:FourldataFit} shows the $\mfourl$ distribution of the data
together with the result of the fit to the \hfourl\ candidates.
The small shape fluctuations in the background prediction arise from the integrations of the background
probability density functions; their effects on the mass fit are negligible.
The fit results in
\begin{equation*}
\mH = \ensuremath{124.99\pm0.18\text{(stat.)}\pm0.04\text{(syst.)}}~\GeV.
\end{equation*}
All fitted normalisations are found to be compatible with the SM expectations.
By allowing the fit to determine the normalization in each subchannel, the result is insensitive to the SM predictions for each subchannel.
There is negligible change in the fit results if the relative fractions are constrained to the SM.
The fit for \mH\ is also performed independently for each decay channel.
The \mfourl\ distributions for the four channels are shown in
Figure~\ref{fig:FourldataFit_all_channels}.
The resulting profile-likelihood-ratio variation as a function of \mH\ and the
resulting fitted values are
compared with those of the combined fit in \Figure{fig:FourlChannels}.
The \mH\ measurements from the individual channels are compatible with the combined measurement,
with a $p$-value of 0.82.

\begin{figure}[htb!]
\centering
\includegraphics[width=0.60\textwidth]{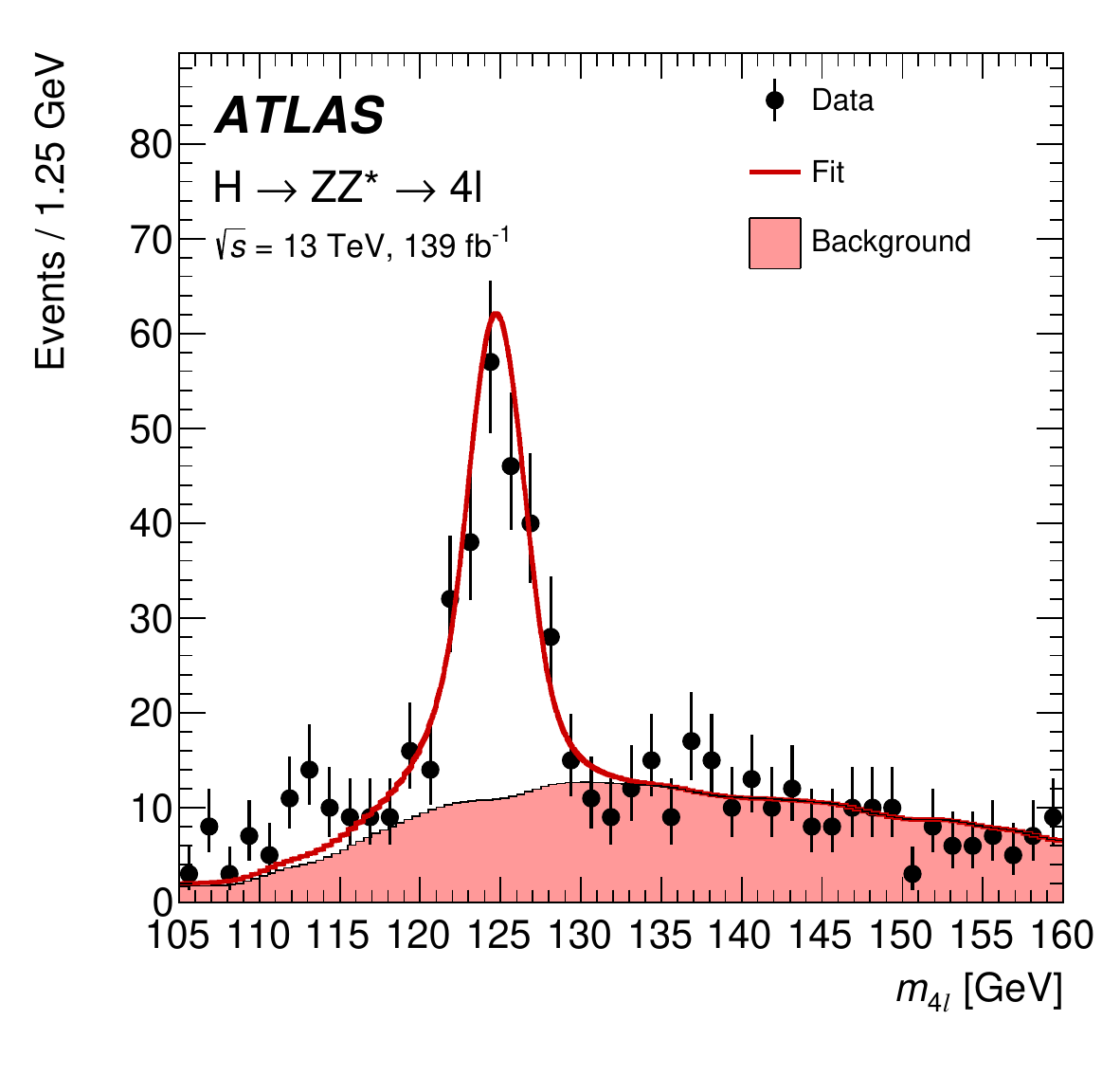}
\caption{The \mfourl\ data distribution from all subchannels combined (black points)
is shown along with the signal-plus-background post-fit probability density function (red line).
The $p$-value quantifying the agreement between the data and the fit model is 0.53.
\label{fig:FourldataFit}}
\end{figure}
 
\begin{figure}[t]
\centering
\subfloat[ ]{\includegraphics[width=0.48\textwidth]{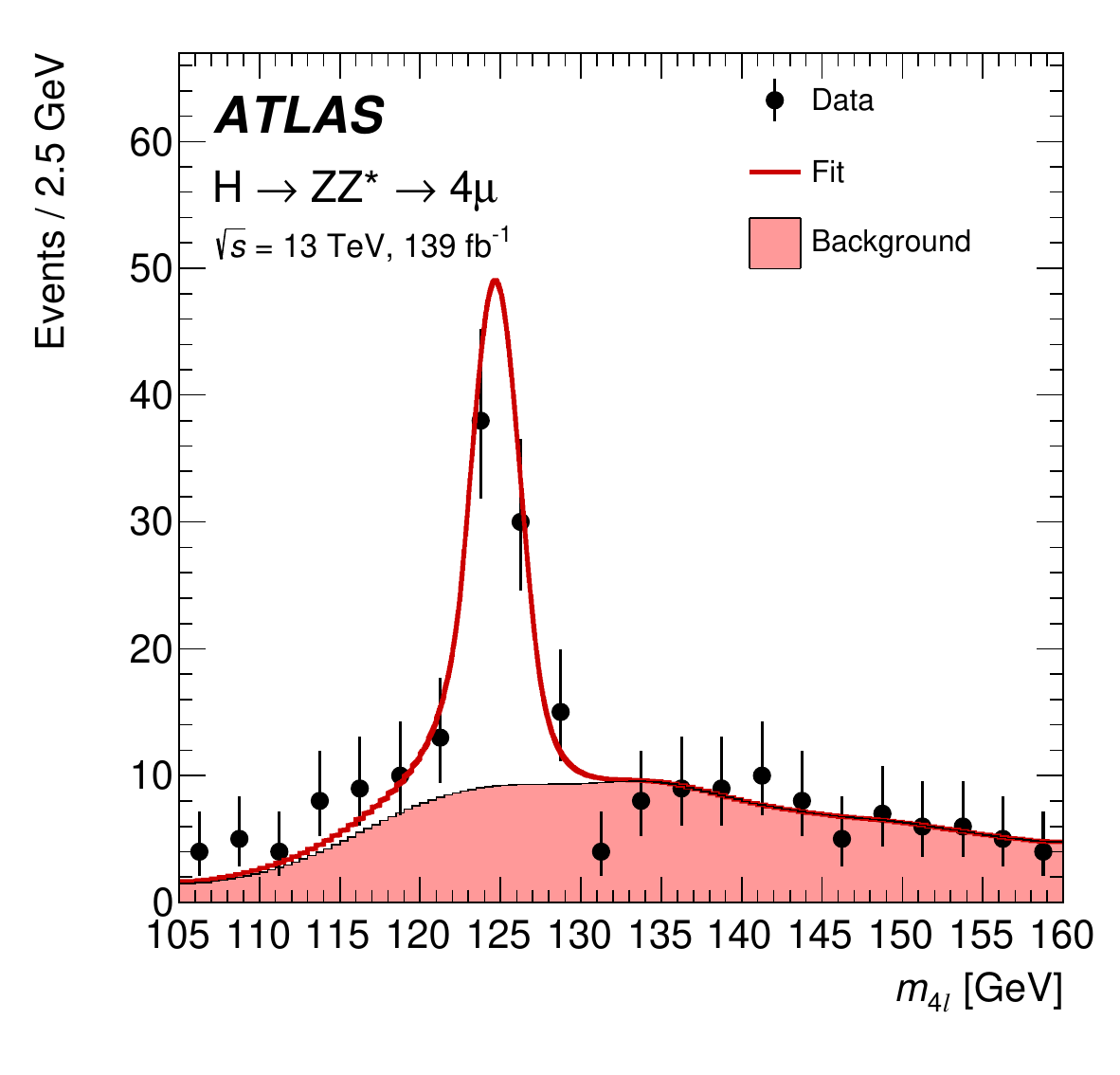}}
\subfloat[ ]{\includegraphics[width=0.48\textwidth]{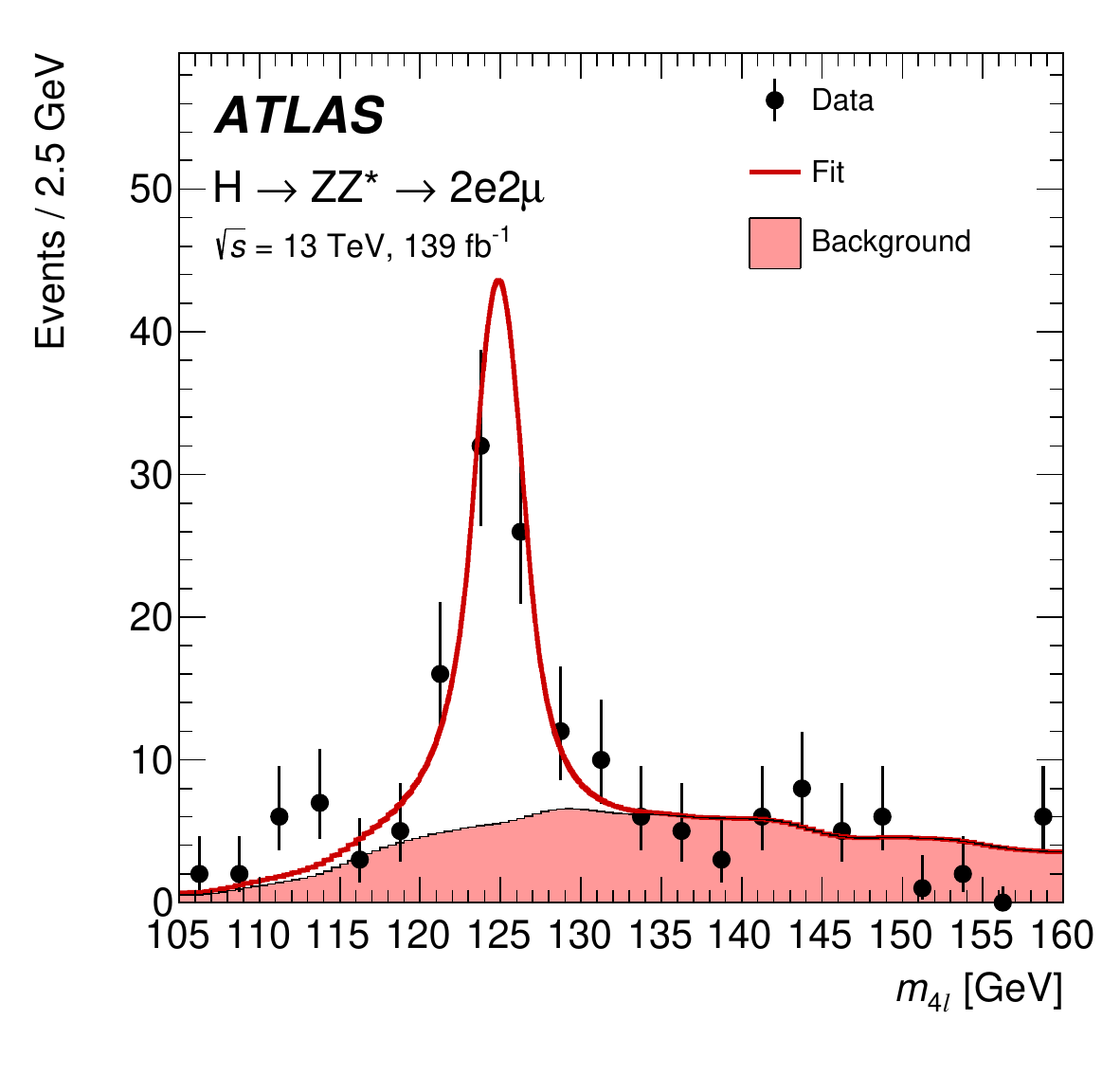}}
 
\subfloat[ ]{\includegraphics[width=0.48\textwidth]{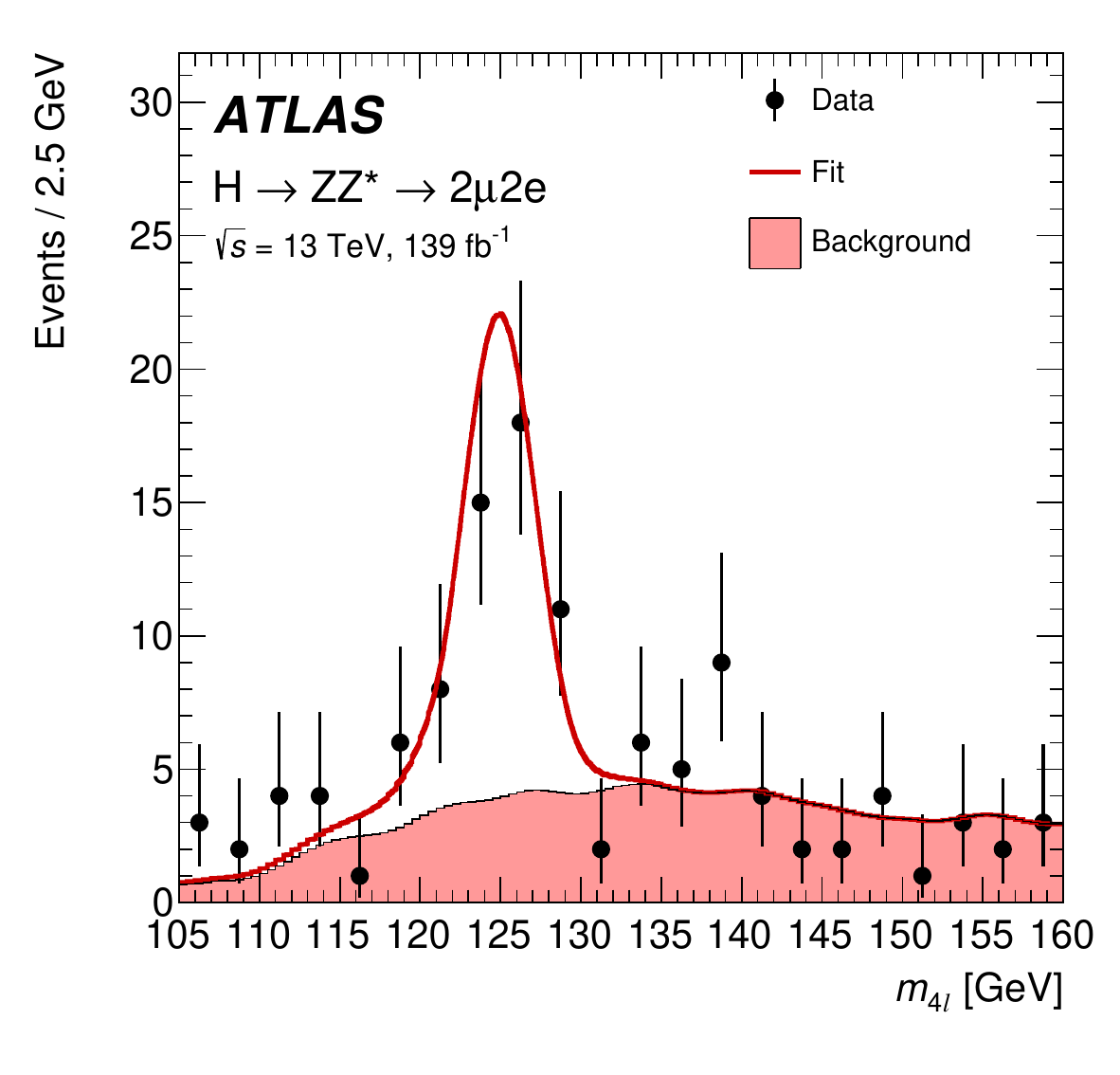}}
\subfloat[ ]{\includegraphics[width=0.48\textwidth]{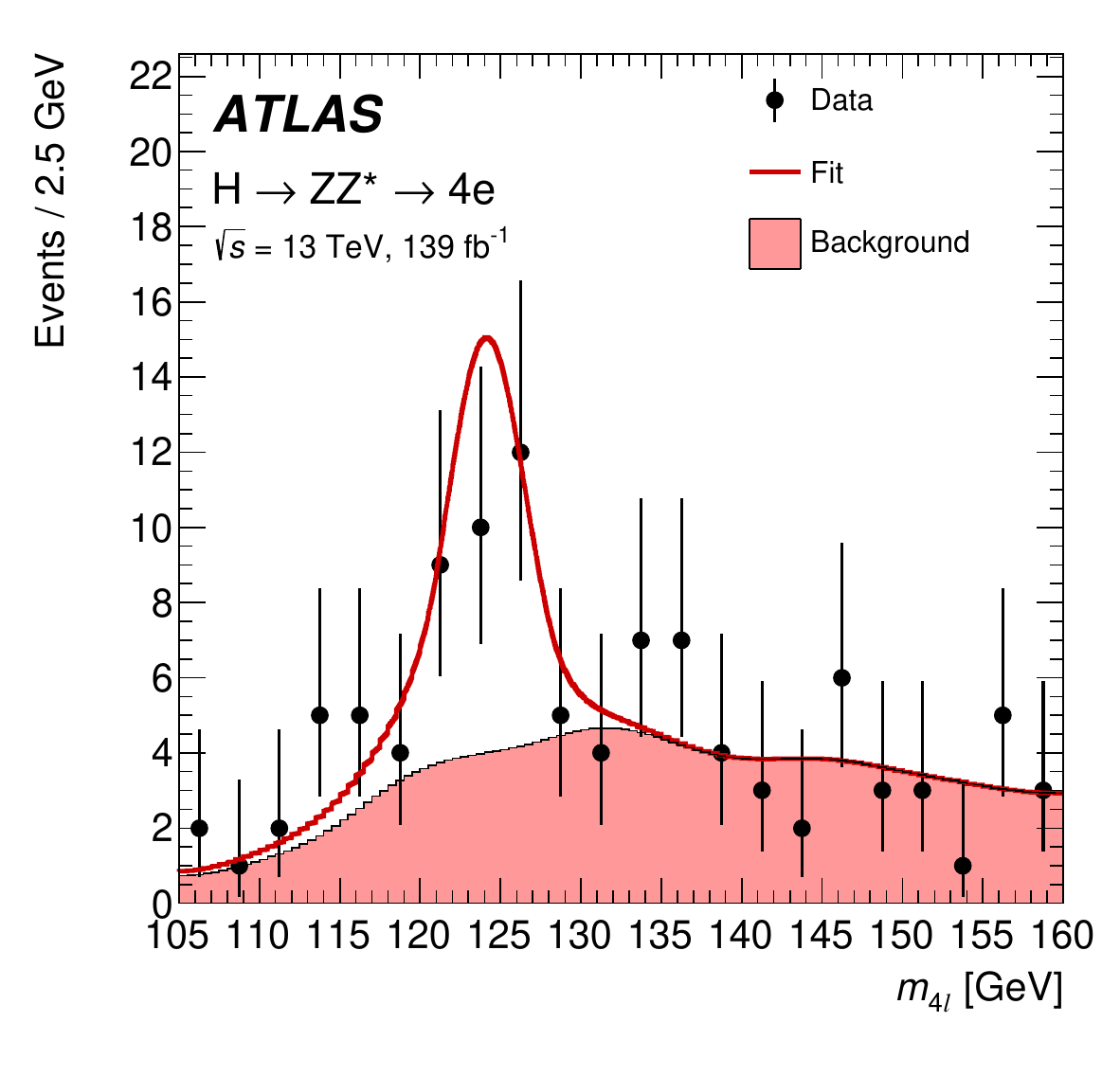}}
\caption{The \mfourl\ data distribution (black points) for
(a) $4\mu$, (b) $2e2\mu$, (c) $2\mu2e$, and (d) $4e$ events
is shown along with the signal-plus-background post-fit probability density function (red line).
The $p$-values quantifying the agreement between the data and the fit model are (a) 0.92, (b) 0.10, (c) 0.83, and (d) 0.97.
}
\label{fig:FourldataFit_all_channels}
\end{figure}
 
\begin{figure}[t]
\centering
\subfloat[ ]{\includegraphics[width=0.47\textwidth]{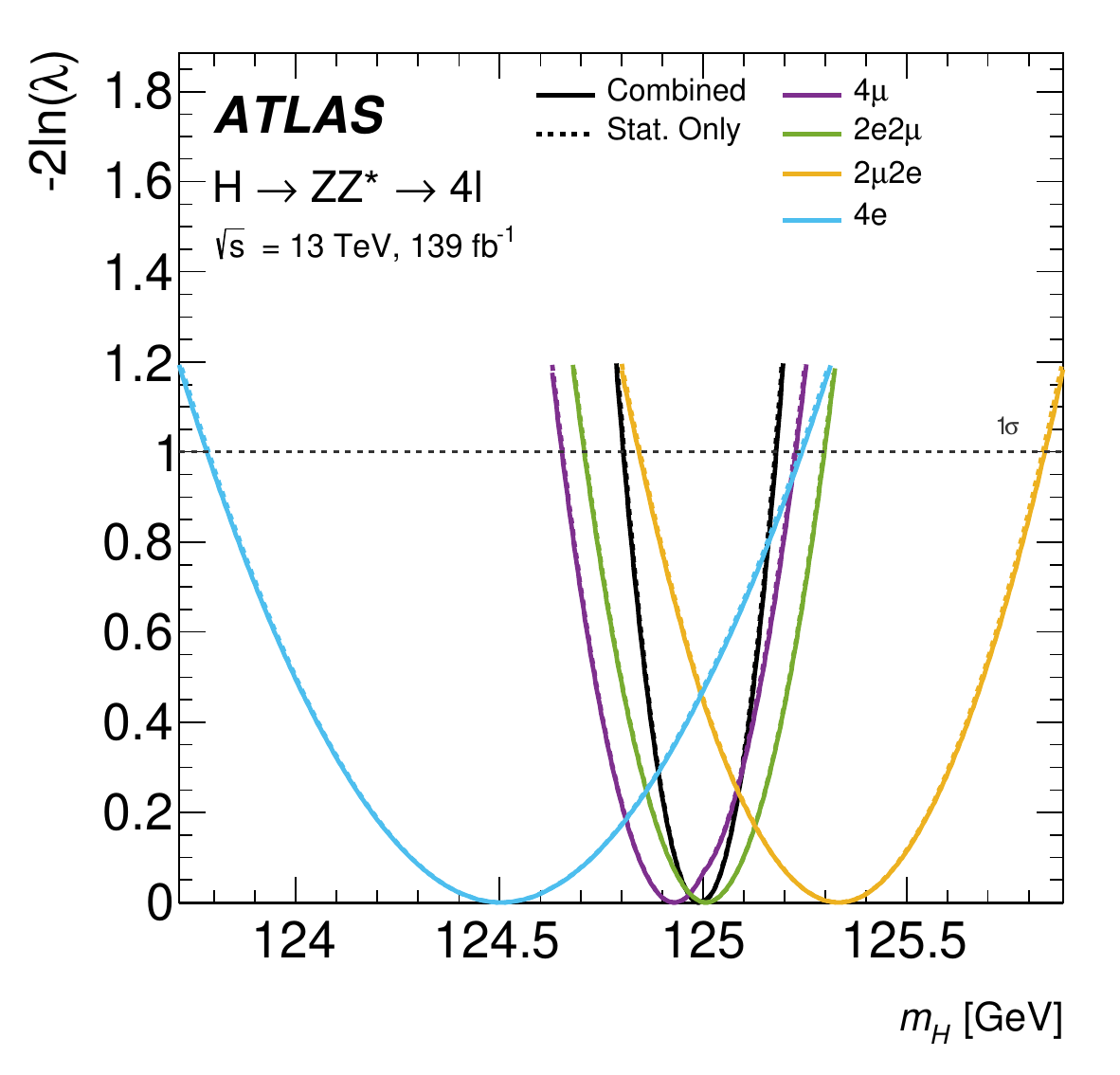}}
\subfloat[ ]{\includegraphics[width=0.55\textwidth]{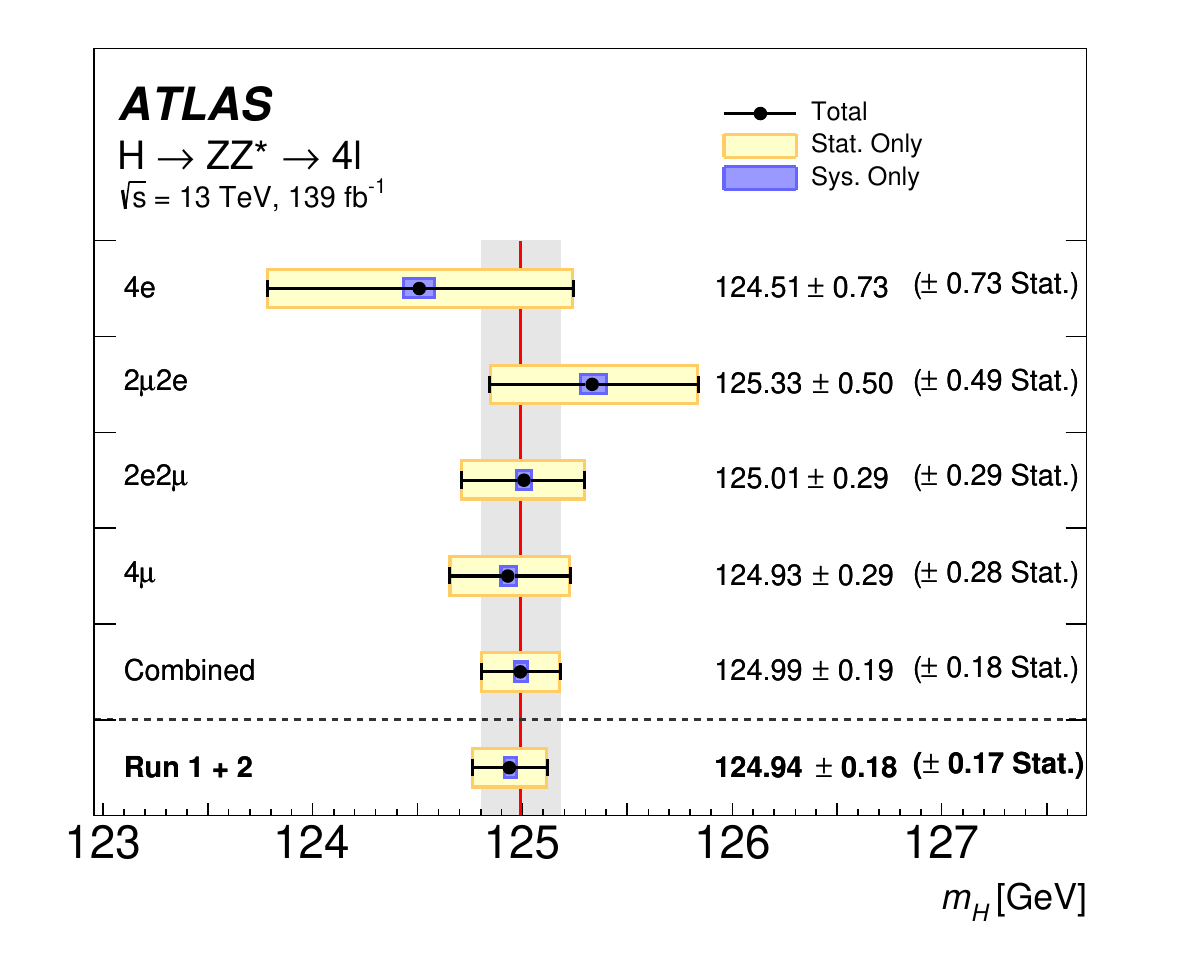}}
\caption{The test statistic, $-2\textrm{ln}(\lambda)$, values as a function of \mH are shown in (a) for the fit in each of the final states
$4\mu$ (purple), $2e2\mu$ (green), $2\mu 2e$ (orange), and $4e$ (blue),
and for the combined fit (black), both with (solid lines) and without (dashed lines)
systematic uncertainties.
The horizontal dashed line
indicates the location of the one-$\sigma$ uncertainty.
The fit results obtained in each final state are shown in (b)
together with the combined result.
The uncertainty bar on each point is the total statistical and systematic uncertainty; the brackets
show the statistical uncertainty only. The vertical (red) line indicates
the combined Run~2 result, and the grey band its total uncertainty.
\label{fig:FourlChannels}}
\end{figure}
 
The statistical uncertainty of \mH\ is estimated by fixing all
nuisance parameters associated with systematic uncertainties to their best-fit values, leaving all remaining parameters
unconstrained.
Uncertainties on the Higgs boson signal originating from the QCD scale, PDF variations, and
modelling of FSR account result in an uncertainty in \mH\ of 14~\MeV.
Other sources, including background modelling and simulation statistics, contribute less than
5~\MeV.
The systematic uncertainties are applied following the strategy detailed in Ref.~\cite{HIGG-2018-28}.
The total systematic uncertainty is estimated by subtracting the square of
the statistical uncertainty from the square of the total uncertainty and calculating the square root.
\Table{tab:syst} shows the leading contributions to the
systematic uncertainty of \mH.
The uncertainties in the electron energy scale and
in the muon momentum scale, resolution, and sagitta bias correction
are described in \Section{sec:objects}.
Other sources, including background modelling and simulation statistics, contribute less than 5~\mev.
The total uncertainty is in agreement with the expectation of $0.19~\GeV$\ and is
dominated by the statistical component.
\Figure{fig:Uncertainty} shows the distribution of the expected uncertainty in \mH\
obtained by fitting a set of pseudo-experiments assuming $\mH = 125$~\GeV.
 
\begin{table}[!htb]
\centering
\caption{Largest contributions to the systematic uncertainty of \mH.
\label{tab:syst}}
\begin{tabular}{l|c}
\toprule
Systematic Uncertainty & Contribution [\mev] \\
\midrule
Muon momentum scale & $\pm 28$ \\
Electron energy scale & $\pm 19$ \\
Signal-process theory & $\pm 14$ \\
\bottomrule
\end{tabular}
\end{table}

\begin{figure}[!htb]
\centering
\includegraphics[width=0.60\textwidth]{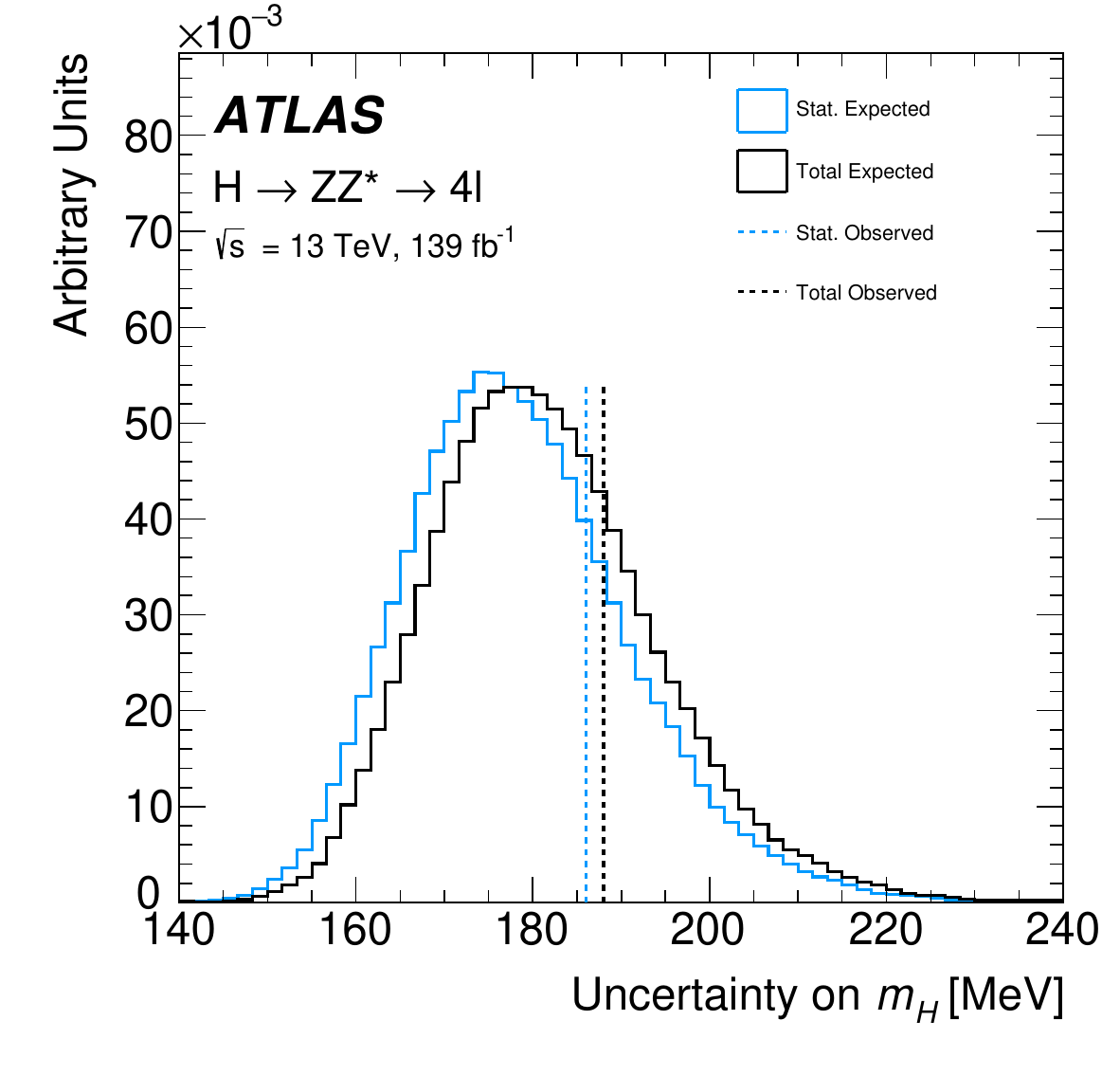}
\caption{The distributions of the total \mH\ uncertainty from
pseudo-experiments assuming $\mH=125$~\GeV\ are shown,
for when the fit does (black) and does not (blue) take into account systematic uncertainties.
The solid lines correspond to the
expected uncertainty distribution from pseudo-experiments,
while the vertical dashed lines indicate the observed values of the uncertainties.
The one-sided $p$-value for compatibility between the observed and expected total uncertainties is 0.28.
\label{fig:Uncertainty}}
\end{figure}
 
The result of this measurement is combined with the measurement of \mH\ using the Run~1 dataset in
the same final states, which was $\mH = 124.51 \pm 0.52$~\cite{HIGG-2013-12}.
The correlation scheme for the systematic uncertainties of the two measurements follows
the one used in Ref.~\cite{HIGG-2016-33},
where only the uncertainties in the lepton calibration were considered correlated.
In this combination, the electron calibration uncertainty is correlated between the two
measurements,
while the muon calibration systematic uncertainty is uncorrelated due to the use of improved
and independent techniques, as described in Section \ref{sec:objects}.
Systematic uncertainties are reduced by about 14\% when using this correlation scheme.
The combined result, calculated by combining the likelihoods of the two measurements is
\begin{equation*}
\mH = \ensuremath{124.94\pm0.17\text{(stat.)}\pm0.03\text{(syst.)}}~\GeV,
\end{equation*}
which has slightly smaller statistical and systematic uncertainties.
The $p$-value for the two individual results, assuming the Higgs boson mass is the combined value, is 0.4.


\FloatBarrier

\section{Summary}
\label{sec:conclusion}
 
The mass of the Higgs boson is measured from a maximum-likelihood fit to the invariant mass
and the predicted invariant-mass resolution of
the \hfourl\ decay channel.
The results are obtained from the full Run~2 $pp$ collision data sample recorded by the ATLAS
experiment at the CERN Large Hadron Collider at a centre-of-mass
energy of 13~\TeV, corresponding to an integrated luminosity
of \intlumi~\ifb.
The measurement is based on the latest calibrations of muons and
electrons, and on improvements to the analysis techniques used
to obtain the previous result using data collected by ATLAS in 2015 and 2016.
 
The measured value of the Higgs boson mass for the \hfourl\ channel is
\[
\mH = \ensuremath{124.99\pm0.18\text{(stat.)}\pm0.04\text{(syst.)}}~\GeV.
\]
Thanks to a larger dataset, improved experimental techniques, and more precise lepton calibration,
the statistical uncertainty of the measurement has been reduced by a factor of two and the
systematic uncertainty by about 20\% relative to the previous Run 2 published result.
 
This measurement is combined with the previous one obtained in the same channel
with ATLAS Run~1 data.
The result of the combination is
\[
\mH = \ensuremath{124.94\pm0.17\text{(stat.)}\pm0.03\text{(syst.)}}~\GeV.
\]
This is the most precise measurement of \mH\ in the \hfourl\ channel
by the ATLAS Collaboration.


\section*{Acknowledgements}


We thank CERN for the very successful operation of the LHC, as well as the
support staff from our institutions without whom ATLAS could not be
operated efficiently.
 
We acknowledge the support of
ANPCyT, Argentina;
YerPhI, Armenia;
ARC, Australia;
BMWFW and FWF, Austria;
ANAS, Azerbaijan;
SSTC, Belarus;
CNPq and FAPESP, Brazil;
NSERC, NRC and CFI, Canada;
CERN;
ANID, Chile;
CAS, MOST and NSFC, China;
Minciencias, Colombia;
MEYS CR, Czech Republic;
DNRF and DNSRC, Denmark;
IN2P3-CNRS and CEA-DRF/IRFU, France;
SRNSFG, Georgia;
BMBF, HGF and MPG, Germany;
GSRI, Greece;
RGC and Hong Kong SAR, China;
ISF and Benoziyo Center, Israel;
INFN, Italy;
MEXT and JSPS, Japan;
CNRST, Morocco;
NWO, Netherlands;
RCN, Norway;
MEiN, Poland;
FCT, Portugal;
MNE/IFA, Romania;
JINR;
MES of Russia and NRC KI, Russian Federation;
MESTD, Serbia;
MSSR, Slovakia;
ARRS and MIZ\v{S}, Slovenia;
DSI/NRF, South Africa;
MICINN, Spain;
SRC and Wallenberg Foundation, Sweden;
SERI, SNSF and Cantons of Bern and Geneva, Switzerland;
MOST, Taiwan;
TAEK, Turkey;
STFC, United Kingdom;
DOE and NSF, United States of America.
In addition, individual groups and members have received support from
BCKDF, CANARIE, Compute Canada and CRC, Canada;
COST, ERC, ERDF, Horizon 2020 and Marie Sk{\l}odowska-Curie Actions, European Union;
Investissements d'Avenir Labex, Investissements d'Avenir Idex and ANR, France;
DFG and AvH Foundation, Germany;
Herakleitos, Thales and Aristeia programmes co-financed by EU-ESF and the Greek NSRF, Greece;
BSF-NSF and GIF, Israel;
Norwegian Financial Mechanism 2014-2021, Norway;
NCN and NAWA, Poland;
La Caixa Banking Foundation, CERCA Programme Generalitat de Catalunya and PROMETEO and GenT Programmes Generalitat Valenciana, Spain;
G\"{o}ran Gustafssons Stiftelse, Sweden;
The Royal Society and Leverhulme Trust, United Kingdom.
 
The crucial computing support from all WLCG partners is acknowledged gratefully, in particular from CERN, the ATLAS Tier-1 facilities at TRIUMF (Canada), NDGF (Denmark, Norway, Sweden), CC-IN2P3 (France), KIT/GridKA (Germany), INFN-CNAF (Italy), NL-T1 (Netherlands), PIC (Spain), ASGC (Taiwan), RAL (UK) and BNL (USA), the Tier-2 facilities worldwide and large non-WLCG resource providers. Major contributors of computing resources are listed in Ref.~\cite{ATL-SOFT-PUB-2021-003}.


\clearpage

\printbibliography
 
\clearpage
 
\begin{flushleft}
\hypersetup{urlcolor=black}
{\Large The ATLAS Collaboration}

\bigskip

\AtlasOrcid[0000-0002-6665-4934]{G.~Aad}$^\textrm{\scriptsize 101}$,
\AtlasOrcid[0000-0002-5888-2734]{B.~Abbott}$^\textrm{\scriptsize 119}$,
\AtlasOrcid[0000-0002-7248-3203]{D.C.~Abbott}$^\textrm{\scriptsize 102}$,
\AtlasOrcid[0000-0002-1002-1652]{K.~Abeling}$^\textrm{\scriptsize 55}$,
\AtlasOrcid[0000-0002-8496-9294]{S.H.~Abidi}$^\textrm{\scriptsize 29}$,
\AtlasOrcid[0000-0002-9987-2292]{A.~Aboulhorma}$^\textrm{\scriptsize 35e}$,
\AtlasOrcid[0000-0001-5329-6640]{H.~Abramowicz}$^\textrm{\scriptsize 150}$,
\AtlasOrcid[0000-0002-1599-2896]{H.~Abreu}$^\textrm{\scriptsize 149}$,
\AtlasOrcid[0000-0003-0403-3697]{Y.~Abulaiti}$^\textrm{\scriptsize 116}$,
\AtlasOrcid[0000-0003-0762-7204]{A.C.~Abusleme~Hoffman}$^\textrm{\scriptsize 136a}$
% ...
\bigskip
\\

% ...
$^{29}$Physics Department, Brookhaven National Laboratory, Upton NY; United States of America.\\
% ...
$^{35}$$^{(a)}$Facult\'e des Sciences Ain Chock, R\'eseau Universitaire de Physique des Hautes Energies - Universit\'e Hassan II, Casablanca;$^{(b)}$Facult\'{e} des Sciences, Universit\'{e} Ibn-Tofail, K\'{e}nitra;$^{(c)}$Facult\'e des Sciences Semlalia, Universit\'e Cadi Ayyad, LPHEA-Marrakech;$^{(d)}$LPMR, Facult\'e des Sciences, Universit\'e Mohamed Premier, Oujda;$^{(e)}$Facult\'e des sciences, Universit\'e Mohammed V, Rabat;$^{(f)}$Institute of Applied Physics, Mohammed VI Polytechnic University, Ben Guerir; Morocco.\\
% ...
$^{55}$II. Physikalisches Institut, Georg-August-Universit\"{a}t G\"ottingen, G\"ottingen; Germany.\\
% ...
$^{101}$CPPM, Aix-Marseille Universit\'e, CNRS/IN2P3, Marseille; France.\\
$^{102}$Department of Physics, University of Massachusetts, Amherst MA; United States of America.\\
% ...
$^{116}$Department of Physics, New York University, New York NY; United States of America.\\
% ...
$^{119}$Homer L. Dodge Department of Physics and Astronomy, University of Oklahoma, Norman OK; United States of America.\\
% ...
$^{136}$$^{(a)}$Departamento de F\'isica, Pontificia Universidad Cat\'olica de Chile, Santiago;$^{(b)}$Millennium Institute for Subatomic physics at high energy frontier (SAPHIR), Santiago;$^{(c)}$Instituto de Investigaci\'on Multidisciplinario en Ciencia y Tecnolog\'ia, y Departamento de F\'isica, Universidad de La Serena;$^{(d)}$Universidad Andres Bello, Department of Physics, Santiago;$^{(e)}$Instituto de Alta Investigaci\'on, Universidad de Tarapac\'a, Arica;$^{(f)}$Departamento de F\'isica, Universidad T\'ecnica Federico Santa Mar\'ia, Valpara\'iso; Chile.\\
% ...
$^{149}$Department of Physics, Technion, Israel Institute of Technology, Haifa; Israel.\\
$^{150}$Raymond and Beverly Sackler School of Physics and Astronomy, Tel Aviv University, Tel Aviv; Israel.\\
% ...
$^{a}$ Also Affiliated with an institute covered by a cooperation agreement with CERN.\\
$^{b}$ Also at An-Najah National University, Nablus; Palestine.\\
$^{c}$ Also at Borough of Manhattan Community College, City University of New York, New York NY; United States of America.\\
$^{d}$ Also at Bruno Kessler Foundation, Trento; Italy.\\
$^{e}$ Also at Center for High Energy Physics, Peking University; China.\\
$^{f}$ Also at Centro Studi e Ricerche Enrico Fermi; Italy.\\
% ...

\end{flushleft}


\end{document}